\documentclass[fleqn,usenatbib]{mnras}
\usepackage[T1]{fontenc}
\usepackage{ae,aecompl}
\usepackage{graphicx}
\usepackage{amsmath}
\usepackage{amssymb}
\usepackage{txfonts}

\title[Reading the tea leaves]
{Reading the tea leaves in the $M_{\rm bh}$-$M_{\rm *,sph}$ and $M_{\rm
    bh}$-$R_{\rm e,sph}$ diagrams: 
dry and gaseous mergers with remnant angular momentum}

\author[Graham \& Sahu]
{
Alister W.\ Graham$^1$\thanks{E-mail: AGraham@swin.edu.au},
Nandini Sahu$^{1,2}$
\\
$^1$ Centre for Astrophysics and Supercomputing, Swinburne University of
Technology, Hawthorn, VIC 3122, Australia\\ 
$^2$ OzGrav-Swinburne, Centre for Astrophysics and Supercomputing, Swinburne
University of Technology, Hawthorn, VIC 3122, Australia
}

\date{Accepted XXX. Received YYY; in original form ZZZ}
\pubyear{2022}

\begin{document}
\label{firstpage}
\pagerange{\pageref{firstpage}--\pageref{lastpage}}
\maketitle

\begin{abstract}

We recently revealed that bulges and elliptical galaxies broadly define distinct, 
super-linear relations in the $M_{\rm bh}$-$M_{\rm *,sph}$ diagram, 
with the order-of-magnitude lower $M_{\rm bh}/M_{\rm *,sph}$ ratios 
in the elliptical galaxies due to major (disc-destroying, elliptical-building)
dry mergers.  Here we present a more nuanced picture.  
Galaxy mergers, in which the net orbital angular momentum does not cancel, 
can lead to systems with a rotating disc.  This situation can occur with either 
wet (gas-rich) mergers involving one or two spiral galaxies, e.g., NGC~5128, 
or dry (relatively gas-poor) collisions involving one or two lenticular galaxies, e.g., NGC~5813. 
The spheroid and disc masses of the progenitor galaxies and 
merger remnant dictate the shift in the $M_{\rm bh}$-$M_{\rm *,sph}$ and
$M_{\rm bh}$-$R_{\rm e,sph}$ diagrams.  We show how this explains  
the (previously excluded merger remnant) S\'ersic S0 galaxies near the bottom
of the elliptical sequence and 
core-S\'ersic S0 galaxies at the top of the bulge sequence, neither of which
are faded spiral galaxies. 
Different evolutionary pathways in the scaling diagrams are discussed.
We also introduce two ellicular (ES) galaxy types, 
explore the location of brightest cluster galaxies 
and stripped `compact elliptical' galaxies in the $M_{\rm bh}$-$M_{\rm
  *,sph}$ diagram, and present a new merger-built $M_{\rm bh}$-$M_{\rm
  *,sph}$ relation which may prove helpful for studies of nanohertz gravitational waves.
This work effectively brings into the fold many systems previously considered outliers with 
either overly massive or undermassive black holes relative to the near-linear 
$M_{\rm bh}$-$M_{\rm *,sph}$ `red sequence' 
patched together with select bulges and elliptical galaxies. 



\end{abstract}


\section{Introduction}

For a
long time, most astronomers, including the author, had, in effect, been looking at the tip of the
proverbial iceberg in the (black hole mass, $M_{\rm bh}$)-(spheroid stellar
mass, $M_{\rm *,sph}$) diagram due to samples 
biased with high fractions of massive red galaxies in which star-formation had
largely ceased.  For many years, the $M_{\rm bh}/M_{\rm *,sph}$ ratio was
routinely quoted as a constant value and used as such in theory, simulations
and observation-based predictions.  This practice was a consequence of belief
in a near-linear $M_{\rm bh}$-$M_{\rm *,sph}$ relation
\citep{1988ApJ...324..701D, 1998AJ....115.2285M} coupled with the accurate
insight and deduction by \citet{1989RvMA....2....1R} that ``massive black
holes now starved of fuel [] lurk in the nuclei of most galaxies''.  However,
a decade ago, this was explained as an incomplete picture, 
and early clues that a universal, near-linear $M_{\rm bh}$-$M_{\rm *,sph}$
relation was not the full story can be found in \citet{1998ApJ...505L..83L},
\citet{2001ApJ...553..677L} and \citet{1999ApJ...519L..39W}. 
The lower-mass spheroids ($M_{\rm *,sph}\lesssim10^{11}$ M$_\odot$), in particular
those without a core partially-depleted of stars, were observed to follow a
near-quadratic $M_{\rm bh}$-$M_{\rm *,sph}$ relation, and dry\footnote{The
  term `dry' denotes systems (galaxies and mergers) relatively poor in 
  star-forming gas, when compared to most spiral galaxies.} mergers were
invoked to explain movement off this relation through the creation of core-S\'ersic
spheroids which appeared to follow a near-linear relation \citep{2012ApJ...746..113G,
  2013ApJ...764..151G, 2013ApJ...768...76S}.  That is, step-change mergers,
discussed by \citet{1980Natur.287..307B} and \citet{1989RvMA....2....1R}, were
heralded as the driving force behind the apparent near-linear $M_{\rm
  bh}$-$M_{\rm *,sph}$ relation at high masses.  

This core-S\'ersic versus
S\'ersic divide was interpreted by \citet{2012ApJ...746..113G} using the
galaxy luminosity-(velocity dispersion) relations, and it was subsequently explored by
\citet[][hereafter KH13, see their Section 6.7]{2013ARA&A..51..511K} and
\citep[][see their Section~5]{2019ApJ...887...10S}.  `Quasar mode', also known
as `cold mode', feedback processes \citep[e.g.,][]{1969Natur.223..690L,
  2000MNRAS.311..576K, 2001ApJ...551..131C, 2010A&A...518L.155F} were considered by
\citet{2013ApJ...764..151G} as responsible for shaping the non-linear
spheroid/(black hole) coevolution at low masses.  Furthermore, the virial masses of
black holes (BHs) in active galactic nuclei (AGNs) were seen to extend this
near-quadratic $M_{\rm bh}$-$M_{\rm *,sph}$ relation down to $M_{\rm
  bh}\approx 10^5$-$10^6$ M$_\odot$ \citep{2015ApJ...798...54G}.  This marked
a significant change, building on past works which had advocated a single
near-linear relation with slope $\sim$0.9--1.2
\citep[e.g.,][KH13]{1995ARA&A..33..581K, 2002MNRAS.331..795M, 2003ApJ...589L..21M,
  2004ApJ...604L..89H, 2007MNRAS.379..711G, 2009ApJ...698..198G,
  2013ApJ...764..184M}.

Over more recent years, our understanding of the $M_{\rm bh}$-$M_{\rm *,sph}$
diagram has advanced in several important ways.  \citet{2016ApJ...817...21S}
observed the presence of galaxies with partially depleted cores at the top of
the near-quadratic relation defined by spheroids without such cores, and they
also revealed how a near-linear ``red sequence'' involving early-type galaxies
(ETGs) yields an apparent separation from the bulges of late-type galaxies (LTGs) which
followed a notably steeper ``blue sequence''.  Doubling the sample size of LTGs
with multicomponent decompositions, \citet{2019ApJ...873...85D} reported that
the LTGs followed a near-quadratic $M_{\rm bh}$-$M_{\rm *,sph}$ relation.
Doubling the sample size of ETGs with similar decompositions,
\citet{2019ApJ...876..155S} discovered that the bulges of lenticular (S0) galaxies
also follow a near-quadratic relation, as does the ensemble of elliptical (E) 
galaxies, albeit with different intercept points, i.e., normalisations.  Indeed, a striking
revelation was the order of magnitude offset in the $M_{\rm bh}/M_{\rm *,sph}$
ratios, at a given spheroid mass, between E galaxies and the bulges
of S0 galaxies.  This offset has since been explained in
\citet{Graham:Sahu:22a} by dry, 
major mergers of lenticular galaxies folding in their disc mass and creating
the elliptical galaxies \citep{2006ApJ...636L..81N}.  
\citet{Graham:Sahu:22a} borrowed from
Darwinian evolution and used the terminology of ``gradualism'' to describe
coevolution via accretion of external material, star formation, AGN growth,
and feedback, and the term ``punctuated equilibrium'' to describe dry mergers
that dramatically change the properties of the remnant galaxy through simple
additive means.\footnote{Technically, the Darwinian analogy may not be perfect
  if accretion-driven black hole growth spurts occur on a shorter timescale
  than the resettlement of disc$+$bulge stars during a galactic merger event.}
While much emphasis and effort has been expended on the role of feedback over
the past quarter century, here, we provide a further, more complete
investigation into these merger-induced jumps in the $M_{\rm bh}$-$M_{\rm
  *,sph}$ diagram.

While it has long been suggested that galaxies with discs can collide to create
an E galaxy \citep[e.g.,][]{1977A&A....54..121V, 1977egsp.conf..401T}, 
they can also merge to create a system with
substantial orbital angular momentum \citep[e.g.,][]{1996ApJ...471..115B,
  2003ApJ...597..893N, 2006ApJ...636L..81N}.  Indeed, the merger simulations by
\citet{1983MNRAS.205.1009N}, 
\citet{1992ApJ...393..484B}, and \citet{2006MNRAS.372..839N} appear quite
capable of creating galaxies in which varying fractions of (initially) disc
stars end up in the spheroidal component of the remnant.  
This outcome seems 
inescapable given the possible range of initial bulge-to-disc ratios, disc
rotation speeds and relative orientations, orbital impact parameters, etc.
In a sense, `natural selection' occurs when the environment and initial
conditions dictate the survival of discs and the creation of spheroids.
Many of the progenitor galaxies' disc stars are at larger orbital radii than
its spheroid stars, and these stars likely contribute to 
the shallow tails of the high-$n$ light profiles of massive ETGs. 

Knowing the mass of the progenitor galaxies' black holes, one can hope to
track the evolutionary paths of merging galaxies through the $M_{\rm
  bh}$-$M_{\rm *,sph}$ diagram.  This task is relatively easy in the case of
dry mergers, where the variable parameter is the fraction of progenitor disc
stars that end up in the new spheroid, related to the net angular momentum of
the pre-merged binary galaxy system. In the case of `wet mergers' involving
cold gas, there will 
additionally be some fraction of star formation and AGN accretion contributing
to the spheroid and black hole mass, respectively
\citep[e.g.,][]{2008MNRAS.391.1137L, 2011ApJ...730....4B,
  2012MNRAS.425.1320I}. 

In wet and dry mergers, there may also be some black hole mass loss through
gravitational radiation \citep[e.g.,][]{1964PhRv..136.1224P,
  1979sgrr.work..191B, 1992PhRvD..45.1013K, 1995ApJ...446..543R,
  2001ApJ...562..297L, 2004ApJ...611..623S, 2020ApJ...897...86C}.  This latter
aspect, or more generally, black hole collisions, adds another exciting
research element beyond the topic of galaxy/(black hole) coevolution and has
led to pioneering endeavours to detect long-wavelength gravitational waves
using `pulsar timing arrays' \citep[e.g.,][]{2020ApJ...905L..34A,
  2021MNRAS.508.4970C, 2021ApJ...917L..19G, 2022ApJ...932L..22G,
  2022arXiv220609289T}.  Most predictions of the expected signal strength of
the stochastic gravitational-wave background --- arising from the merger of
many supermassive black hole binaries --- at nanohertz-frequencies have been
based on the notion of a near-linear $M_{\rm bh}$-$M_{\rm *,sph}$ relation.
However, \citet{Graham:Sahu:22a} revealed that one might be better served
using the super-linear $M_{\rm bh}$-$M_{\rm *,sph}$ relation defined by
merger-built elliptical (and Es,e: see Section~\ref{Sec_ESe-b})
galaxies.\footnote{A more complete approach, which we provide here, would also
  offer a prescription for including the (wet and dry) merger-built S0
  galaxies.}

In this work, we present a value-added $M_{\rm bh}$-$M_{\rm *,sph}$ diagram, 
which better reveals the phylogeny of spheroids and shows the (possible origin
and) location of different types of merger remnant.
%
%
We pay particular attention to `mergers' previously flagged for exclusion from
the $M_{\rm bh}$-$M_{\rm *,sph}$ diagram due to their semi-unrelaxed state
and thus potentially evolving spheroid (and black hole) mass.  These include 
lenticular galaxies undergoing star formation, likely involving at least one
spiral galaxy progenitor \citep[e.g.,][]{1977egsp.conf..401T}, and are 
discussed in Section~\ref{Sec_wet_merger}.  These merger-built disc galaxies
follow the low-mass end of the elliptical galaxy $M_{\rm
  bh}$-$M_{\rm *,sph}$ relation.  Pseudobulges, sometimes claimed as the
population at the low-mass end of the spiral galaxy sequence
\citep[e.g.,][]{1996ApJ...457L..73C,2001ApJ...546..216C}, are discussed in
subsection~\ref{Sec_pseudo}.  In Section~\ref{Sec_dry_merger}, we address
lenticular galaxies whose bulges have partially depleted cores, suggestive
that these galaxies were built from a dry merger event in which the net
angular momentum did not cancel.  We also discuss 
the ellicular (ES) galaxy type and potential evidence for subtypes
differentiating between merger remnants midway between an E and an S0 galaxy
versus compact relic `red nuggets' \citep{2005ApJ...626..680D,
  2011ApJ...739L..44D} that have accreted an intermediate-scale disc.  We also
review the location of the brightest cluster galaxies (BCGs:
Section~\ref{Sec_BCGs}) in the $M_{\rm bh}$-$M_{\rm *,sph}$ diagram.
Finally, in Section~\ref{Sec_Strip}, we note how the stripping of stars by the
gravitational tide of a massive neighbouring galaxy can impact the placement
of systems such as `compact elliptical' (cE) galaxies and `ultracompact dwarf
galaxies' (UDGs) in the $M_{\rm bh}$-$M_{\rm *,sph}$ diagram.  A brief
summary is provided in Section~\ref{Sec_Sum}, prior to an Appendix which
provides new multicomponent decompositions for seven galaxies and possibly
helpful notes on several others.

\section{The Data Sample}\label{Sec_data}

This research forms an extension of \citet{Graham:Sahu:22a}, which was based
on a sample of 104 galaxies with directly measured supermassive black hole
masses, critical galaxy morphology information, and multicomponent decompositions
of their Spitzer Space Telescope images taken at an infrared wavelength of
3.6~$\mu$m.  Of these 104, 31 are late-type galaxies (S), 35 are elliptical
galaxies, and the remaining 38 are identified as S0 or ES
galaxies.\footnote{\citet{1966ApJ...146...28L} introduced the ES galaxy type
  discussed further in \citet{2019MNRAS.487.4995G}.}  This paper uses the same
sample, black hole masses, spheroid and galaxy stellar masses, and spheroid
sizes, but looks more deeply into this rich data set.  For instance, some of the S0
galaxies are clearly built from wet mergers and others from dry mergers.
That is, they are not all faded S galaxies.  More generally, a greater
emphasis is placed on the subtleties of the ETG morphology and distinguishing
between E, ES and S0 galaxies and whether the latter two types have been
built by mergers.

Rather than simple bulge$+$disc fits, which can fail to capture the
bulge component correctly, or randomly fitting multiple (e.g., three)
\citet{1963BAAA....6...41S} $R^{1/n}$ functions
that may have no connection to the physical structures in galaxies, the images
were previously inspected by us for the presence of distinct components.
Details of this process were noted in \citet{Graham:Sahu:22a}, see also
\citet{2015ApJ...810..120C}, and the decompositions are shown for each galaxy
in \citet{2016ApJS..222...10S}\footnote{Readers may find the paper's original format, available
  at arxiv.org, preferable.}, \citet{2019ApJ...876..155S}\footnote{See the 
  corresponding journal's online link ``View figure set (41 panels)'' after
  their Fig.~3.} and \citet{2019ApJ...873...85D}.

One point worth noting is that our image analysis will often conveniently fold
boxy/X/(peanut shell)-shaped structures --- captured with radially-dependent,
Fourier harmonic 
terms used to describe isophotal departures from pure ellipses
\citep[e.g.,][]{2016MNRAS.459.1276C} --- back into the bar from which they 
emerged \citep[e.g.,][]{1990A&A...233...82C, 2016ASSL..418..391A}.  
When this does not occur, we add a barlens component to the decomposition. 
As such, for the most part, our bulge luminosities and masses do not pertain to such
`pseudobulges'.  Nor do our bulge masses pertain to bars or inner discs, which
we model as separate components when present.  Based on a multitude of clues
from our past work, for seven galaxies, we reinvestigated their image 
and provide an updated decomposition in Appendix~\ref{Apdx1}.
We remodel four of the original galaxies from
\citet{2016ApJS..222...10S}, two ETGs from \citet{2019ApJ...876..155S}, and
one spiral galaxy from \citet{2019ApJ...873...85D}.

Our spheroid stellar masses were based on the 3.6~$\mu$m magnitudes and the
colour-dependent, spatially-constant, stellar mass-to-light ratio, $\Upsilon_*$, prescription from
the dusty galaxy model of \citet[][their Table~6]{2013MNRAS.430.2715I} after
conversion to a \citep{2002Sci...295...82K} stellar initial mass function
\citep[][their Equation~4 and Fig.~1]{Graham:Sahu:22a}.  These 3.6~$\mu$m $\Upsilon_*$
ratios do not vary too much with optical colour, with $0.6 \lesssim \Upsilon_* \lesssim 0.9$ for 
the ETG colours $0.8 < B-V < 1.0$, and providing enhanced stability over
estimates of the bulge (and galaxy) stellar mass derived from optical
magnitudes alone.  As shown in the Appendix of \citet{Graham:Sahu:22a}, a
broadly consistent set of stellar masses was obtained using the independent
set of colour-dependent $\Upsilon_*^{3.6}$ ratios from 
\citet{2022AJ....163..154S}, once converted to the same
\citep{2002Sci...295...82K} stellar initial mass function \citep[][their
  Equation~A2]{Graham:Sahu:22a}.


\subsection{Advancing the $M_{\rm bh}$-$M_{\rm *,sph}$ diagram}

Two contributions we have strived to make over the
years are improved measurements of the spheroid mass coupled with a better recognition
of the host galaxy morphology.  This has been achieved through several
advancements in galaxy image analysis, such as the introduction of the
core-S\'ersic model \citep{2003AJ....125.2951G}, which unites a spheroid's
inner and outer structure.  
\citet{1966ApJ...143.1002K} and \citet{1972IAUS...44...87K} had reported that
massive ETGs have cores which flatten relative to the outer
\citet{1948AnAp...11..247D} $R^{1/4}$ 
light profile.  \citet{1980Natur.287..307B} and \citet{1991Natur.354..212E} explained how binary black holes
from galaxy merger events could create such cores, with massive dense infalling star
systems creating larger cores \cite[e.g.,][]{2001ApJ...560..636E, 2010ApJ...725.1707G,
2016ApJ...829...81B}. 
Building on the `core-Hubble' model of \citet{1993AJ....106.1371C}, and the
double power-law models of \citet{1994AJ....108.1598F} and
\citet{1994AJ....108..102G}, application of the core-S\'ersic model revealed that roughly 1-in-5 past
allegations of a partially depleted core --- when based on the Nuker
model\footnote{The Nuker model is mathematically equivalent to the double
  power-law introduced by \citet[][his Equation~43]{1990ApJ...356..359H}.} 
\citep{1994AJ....108..102G, 1994ESOC...49..147K}, 
--- were not correct
\citep[see][their Appendix~A.2]{2012ApJ...755..163D,
  2013ApJ...768...36D}.\footnote{This was in part due to the measured Nuker model
  parameters depending on the fitted radial range \citep{2003AJ....125.2951G}, but also because 
 low-mass spheroids with low S\'ersic indices have shallow inner light 
 profiles but no depleted core \citet[][their Fig.~8]{2003AJ....125.2936G}. 
The presence of additional nuclear discs 
 can also flatten the inner light profile, creating the illusion of a central
 deficit \citet{2013ApJ...768...36D}.} 
Fitting a S\'ersic model to a spheroid containing a core-S\'ersic light
profile can result in a misleadingly low S\'ersic index and erroneously faint spheroid
luminosity.  The core-S\'ersic model is available in {\sc Profiler}
\citep{2016PASA...33...62C} and was implemented into 
{\sc Galfit} \citep{2010AJ....139.2097P}
by \citet{2014PASP..126..935B} and can 
be found in {\sc Imfit} \citep{2015ApJ...799..226E}.

Coupled 
with the above has been the realisation that the less luminous ($M_B \gtrsim
-20.5\pm0.75$, or $M_V \gtrsim -21.4\pm0.75$ Vega mag) galaxies often extra
lighave `extra light', 
such as a nuclear star cluster, 
rather than a central deficit of light. Furthermore, the 
nuclear star cluster mass has been discovered to scale with the host 
spheroid mass \citep{2003ApJ...582L..79B, 2003AJ....125.2936G}, and nuclear
discs are also a common feature \citep[e.g.,][]{1991A&A...244L..25N, 1994AJ....108.1567J, 2007ApJ...665.1084B}.  
While nuclear star clusters are typically only a fraction of one per cent of the host 
spheroid's stellar mass 
\citep[e.g.,][]{2006ApJ...644L..21F, 2006ApJ...644L..17W, 2013ApJ...763...76S}, 
if they are overlooked in the modelling of the galaxy light, then 
they can have a disproportionate influence on the galaxy decomposition by, for
example, inflating both the fitted S\'ersic index and luminosity of the
spheroid.\footnote{This situation
  arises when sufficiently high spatial resolution reveals an excess in the
  image capable of biasing the galaxy decomposition.}  As such, accounting 
for these additional components in the image analysis can be 
important for acquiring an unbiased spheroid magnitude, and also for better 
understanding the nature of, and connections between, galaxies.  

\begin{figure*}
\begin{center}
\includegraphics[trim=0.0cm 0cm 0.0cm 0cm, width=1.0\textwidth, angle=0]{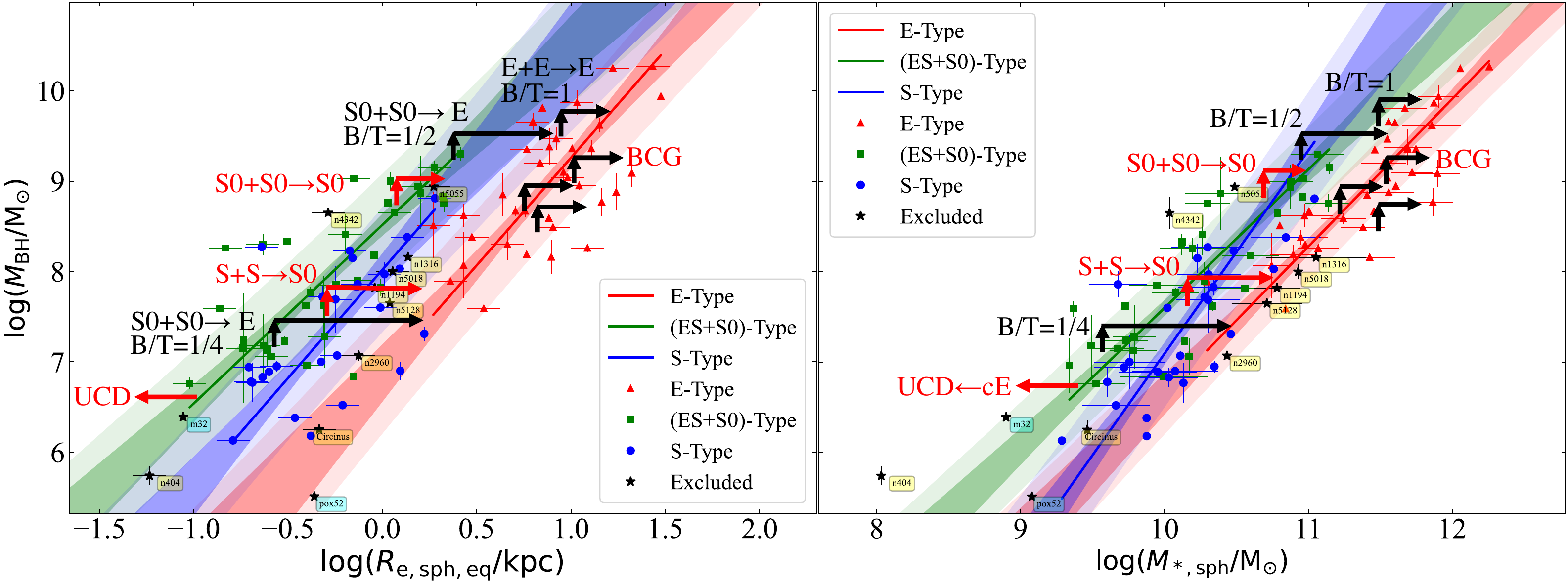}
\caption{ Adaption of Fig.~8 from
  \citet[][using their morphological type]{Graham:Sahu:22a}, where the black
  arrows/tracks denote evolutionary pathways arising from dry, equal-mass
  mergers of ETGs (with bulge-to-total stellar mass ratios as indicated)
  building E galaxies. These growth channels implicitly assume the net
  cancellation of angular momentum from the progenitor galaxies and their
  binary system.  $B/T=1$ mergers encapsulate major E+E collisions building
  new E galaxies and/or BCGs, associated with a doubling of the black hole
  mass and the galaxy mass, while E galaxies built from equal-mass collisions
  of S0 galaxies experience a
  greater increase in spheroid mass depending on the S0 galaxies' initial $B/T$
  ratio.  The red tracks are necessarily more representative in nature, 
   rather than quantitative like the black tracks. 
  The upper red track represents a dry, equal-mass 
  S0 galaxy merger in which the black hole mass doubles and some of the
  angular momentum does not cancel: the 
  newly-built system could be a core-S\'ersic S0 galaxy.  The middle red
  track shows a similar situation but for a wet merger involving spiral
  galaxies creating an S0 galaxy with a bulge likely to have a S\'ersic light
  profile.  A cluster of four such galaxies (NGC~1194, NGC~1316, NGC~5018, and
  NGC~5128) are labelled, along with the Sa
  merger NGC~2960.  They, and the other galaxies with yellow labels were
  excluded from the regression analysis in \citet{Graham:Sahu:22a}, as were
  the galaxies with blue labels for which Spitzer Space Telescope imaging data was not available. 
  This merger driven march to higher masses can place spiral
  galaxy mergers at the lower end of the elliptical galaxy sequence even
  without star formation or black hole accretion.  The lower red track
  reflects tidal stripping of galaxies, capable of shifting a system to the left if
  bulge stars are stripped, while removal of only disc stars --- likely
  explaining some compact elliptical galaxies like M32 --- should yield no
  shift in the $M_{\rm bh}$-$M_{\rm *,sph}$ diagram.  NGC~4342 and possibly
  NGC~4486B (shown later) represent higher-mass S0s that have been stripped.}
\label{Fig-M-R}
\end{center}
\end{figure*}

For example,
\citet[][p.\ 1783]{1997AJ....114.1771F} partly advocated against a division
among ETGs at $M_B\sim -20.5$ mag by arguing that the additional brightness 
of the nuclear star clusters yields a continuous and unifying galaxy 
luminosity-(central surface brightness) 
relation for ETGs brighter than $M_B \sim -18$ mag \citep[][their
  Fig.4c and 6]{1997AJ....114.1771F}.  This followed 
\citet{1985ApJ...295...73K}, which concluded that ETGs brighter and fainter
than  $M_B \sim -18$ mag formed in very different ways. 
Why that interpretation of ETGs was misleading is 
explained in \citet{2003AJ....125.2936G} and detailed further in
\citet{2019PASA...36...35G}, both of which advocated for a divide at
$M_B\sim -20.5$ mag, in agreement with other arguments in 
\citep{1997AJ....114.1771F}.  This divide is mirrored by the (cold
gas)-poor S0  versus E galaxy separation at $\log(M_{*,sph}/M_\odot) \approx 10.9\pm0.2$, seen in
Figure~\ref{Fig-M-R}. It arises from the (large scale disc)-eroding,
spheroid-building, core-depleting mergers of disc galaxies with massive black
holes \citep{1983MNRAS.205.1009N, 1989A&A...215..266N, 1991Natur.354..212E}.

An additional breakthrough of sorts has been the
rediscovery of ES galaxies \citep{2016MNRAS.457..320S}.  Introduced by
\citet{1966ApJ...146...28L} and reviewed by \citet{2019MNRAS.487.4995G}, they
contain intermediate-scale discs having sizes greater than nuclear discs
($\sim$20~pc to a few hundred parsecs) and smaller than the large-scale discs
of S0 galaxies (with typical disc scalelengths of a few kpc). Some of these ES
galaxies are also known as `disc ellipticals' \citep{1988A&A...195L...1N}. 
In
Section~\ref{Sec_ESe-b}, we have introduced the ES,e and ES,b subtypes.
Failing to identify such discs during the decomposition of the galaxy light
can result in an erroneous luminosity for the spheroid. 


Another advance has stemmed from the realisation that the {\sc ELLIPSE} task
\citep{1987MNRAS.226..747J} in the IRAF software suite is flawed, failing to
properly capture the isophotal deviations from pure ellipses observed at least
since \citet{1940ApJ....91..273O} in the case of NGC~3115, and described using
Fourier harmonics by \citet{1978MNRAS.182..797C}, see also
\citet{1987ApJ...312..514C} and \citet{1987A&A...177...71B}.
\citet{2015ApJ...810..120C} revealed how the size of the error in the {\sc
  ELLIPSE} task grows with the ellipticity of the isophote, which means it is
particularly prominent when a bar is present, or a disc is approaching an
edge-on orientation.  Correcting this, \citet{2016MNRAS.459.1276C} were, for
the first time, able to detect and quantify the radial variation in the
amplitude of the sixth order Fourier Harmonic term, which does well with
capturing (peanut shell)-shaped structures 
\citep[e.g.,][]{1974IAUS...58..335D, 1986AJ.....91...65J, 1987MNRAS.229..691S} 
associated with buckled bars 
\citep{1990A&A...233...82C, 2005MNRAS.358.1477A}.  As dramatically illustrated by
\citet{2015ApJ...810..120C}, in the case of NGC~128, the superposition of the
bar and (peanut shell)/box/X-shaped `bulge' results in isophotes which are not
boxy, i.e., neither described by nor captured with the fourth-order Fourier
Harmonic term but instead requiring the sixth order Fourier Harmonic term.
Collectively, this has enabled progress in dissecting and understanding the
$M_{\rm bh}$-$M_{\rm *,sph}$ diagram.  

In this work, we provide some further notes on
individual galaxies, which are common to us and KH13, but for which we
disagree on the spheroid mass. 
Finally, it is noted that there is of course room for further improvement,
which could include application of galaxy-component-specific stellar
mass-to-light ratios for calculation of galaxy component masses, 
and also for the derivation of black hole masses, taking into account the
nuclear star clusters. \citet{2017ApJ...836..237N}
  represents a good example where the nuclear stellar populations in NGC~404 were
  modelled and a spatially-varying colour-dependent $\Upsilon_*$ 
  used.\footnote{\citet{2020MNRAS.496.4061D} subsequently updated the measured
    black hole mass in NGC~404, which we have used.} 
Working with images taken at 3.6~$\mu$m, the influence of hot young stars is
small given that their blackbody radiation curves peak in the blue to UV
region of the electromagnetic spectrum. Instead, older stars, and to a lesser
degree thermally-glowing dust \citep{2015ApJS..219....5Q}, contribute 
to the light seen at 3.6~$\mu$m. 
An analysis of Hubble Space Telescope (HST) images, coupled with
stellar mass-to-light ratios for the nuclear star clusters, plus the refined
galaxy morphologies used here, would enable advances with the $M_{\rm
  nsc}$-$M_*$ relation for LTGs \citep{2003ApJ...582L..79B} and ETGs
\citep{2003AJ....125.2936G}, and with the $(M_{\rm nsc}+M_{\rm bh})$-$M_*$
relation \citep{2009MNRAS.397.2148G} and the $(M_{\rm nsc}+M_{\rm
  bh})$-$\sigma$ relation \citep{2012MNRAS.422.1586G}.  To date, these
discoveries have taken a back seat to the $M_{\rm bh}$-$M_*$ and $M_{\rm
  bh}$-$\sigma$ relations \citep{1998AJ....115.2285M, 2000ApJ...539L...9F,
  2000ApJ...539L..13G}, yet such mass-mass-morphology (m$^3$) and
mass-sigma--morphology (msm) diagrams involving nuclear star clusters will
enable a better understanding of massive compact objects at the centres of
galaxies, as will the $M_{\rm bh}$-$M_{\rm nsc}$ relation
\citep{2016IAUS..312..269G}. HST-resolved images, coupled with the galaxy
morphologies, will also enable an exploration of the depleted cores built by
merging black hole binaries \citep{1980Natur.287..307B, 2005LRR.....8....8M},
and the prior destruction of the nuclear star clusters \citep{2010ApJ...714L.313B}.

\section{Demographics in the (black hole)-spheroid mass diagram}

Within the $M_{\rm bh}$-$M_{\rm *,sph}$ diagram, 
\citet{2019ApJ...876..155S} and 
\citet{Graham:Sahu:22a} present a super-linear bulge sequence, and an offset parallel
elliptical galaxy sequence 
having $M_{\rm bh}/M_{\rm *,sph}$ ratios that differ by roughly an order of
magnitude at any given spheroid mass.  This primary pattern
is not due to splitting off the disc's stellar mass from a single unified 
$M_{\rm bh}$-$M_{\rm *,gal}$ relation. It can, however, 
be explained in terms of elliptical galaxy 
formation via dry S0 galaxy mergers, which end in little net rotation. 

Here, we identify and explain additional substructures in the 
$M_{\rm bh}$-$M_{\rm *,sph}$ diagram, typically when the merger remnant
is not an elliptical galaxy but contains significant rotation, as in the case
of lenticular galaxies. 
We also broach the location of BCGs --- typically E galaxies likely built from
multiple mergers --- in the $M_{\rm bh}$-$M_{\rm *,sph}$ diagram and the
effect of `tidal stripping' on the location of bulges.  These latter two
processes can broaden the distribution in the $M_{\rm bh}$-$M_{\rm *,sph}$
diagram, acting in opposing directions.

\subsection{Bulges at the bottom of the E galaxy $M_{\rm bh}$-$M_{\rm *,sph}$ sequence}
\label{Sec_wet_merger}

KH13 had a subsection titled ``Mergers in progress have abnormally small BH
masses''.  They were referring to the known result that, relative to
the original near-linear $M_{\rm bh}$-$M_{\rm sph}$ relation, 
the $M_{\rm bh}/M_{\rm sph}$ ratio is low in the mergers they mentioned. 
Here, we explain that these system's black hole masses are not abnormal but instead as 
expected, given their origin from a wet galaxy merger likely involving one or
two spiral galaxies. 

KH13 advocated excluding five merger products (NGC~1316, NGC~2960,
NGC~4382, NGC~5128 and IC~1481) 
from the black hole scaling diagrams (see their Table~2). 
These galaxies contain merger-built `classical bulges' whose inclusion, as we
shall see, provides a more complete understanding of galaxy/(black hole) coevolution. 
KH13 regarded these systems as elliptical galaxies; see the final paragraph of their
Section~5, and in their $M_{\rm bh}$-$M_{\rm sph}$ diagram (their Fig.~14), they appear 
as notable outliers.  However, their offset location is somewhat 
exaggerated because the galaxy 
rather than the bulge luminosity was used for these (three S0,pec plus two S?) galaxies. 

While still perturbed at varying levels, these are recognised disc galaxies
whose light profiles display an unmistakable bulge plus outer exponential structure
\citep{2016ApJS..222...10S, 2019ApJ...876..155S} and for which rotation has
been measured. Indeed, in the case of NGC~5128, it has long been known to
rotate \citep{1959ApJ...129..271B, 1995ApJ...449..592H}.  As noted in the
Introduction, wet mergers 
involving spiral galaxies can lead to disc galaxies with more prominent bulges
\citep[e.g.,][]{1996ApJ...471..115B}.  NGC~5128 (aka Centaurus~A) 
is a recognised merger involving at least one spiral galaxy
\citep{1980ApJ...241..969T, 2003MNRAS.338..587B, 2020MNRAS.498.2766W}, as is
NGC~1316 \citep{2021JApA...42...34V}, which underwent a major gas-rich merger
3~Gyr ago \citep[e.g.,][]{2004ApJ...613L.121G}.  NGC~2960 (Mrk~1419) is an
optically disturbed S0--Sa LINER with no nearby companions
\citep{2008ApJ...679.1047K} and labelled a merger product by KH13.
NGC~2960 may represent an advanced stage of the suspected spiral-spiral
merger NGC~7252 \citep{1993AJ....106.1354W}. 

In passing, we note that pushing observations into the low surface brightness
regime around the outskirts of galaxies \citep[e.g.,][]{2001ApJ...549L.199M, 2008ApJ...677..846B, 
2018ApJ...857..144H, 
2022MNRAS.513.1459M}, may be insightful if tidal streams, debris 
tails, shells and ripples \citep[e.g., NGC3923:][]{1986ApJ...306..110F} 
can distinguish between accretion versus more substantial mergers.  As such, deep
observations of the sample of galaxies with directly measured black hole
masses, coupled with their location in the $M_{\rm bh}$-$M_{\rm sph}$
diagram, may reveal further tell-tale clues as to the origin and evolution of galaxies
more generally. 
For example, 
\citet{2018A&A...614A.143M} have flagged the dusty spiral galaxy NGC~3227 as a
possible major merger remnant, somewhat akin to NGC~2960 but with $M_{\rm bh}$ eight
times larger.  Future gains in this exciting area are expected from 
ESA's new ARRAKIHS (Analysis of Resolved Remnants of Accreted galaxies as a Key
Instrument for Halo Surveys) F-class mission, comprised of a dedicated satellite
providing incredibly deep optical and near-infrared images of $\sim$100 nearby galaxies
and their outskirts (P.I. R.Guzm\'an).



While NGC~4382 and IC~1481 are not in our Spitzer sample,
\citet{2019ApJ...887...10S} reported an additional two S0
merger remnants among our Spitzer sample of galaxies with directly measured black hole
masses.  These are NGC~1194 and NGC~5018, an S0 galaxy whose black hole mass
was only recently acquired by \citet{2016ApJ...818...47S}.
\citet{2019ApJ...887...10S} commented on a faint debris tail in NGC~1194, a
galaxy which displays a prominent dust lane somewhat reminiscent of that seen
in NGC~1316 and NGC~5128.  \citet{2016AN....337...96F} explored but could not
confirm the presence of two BHs in this galaxy.  In NGC~5018, the
brightest galaxy within the NGC~5018 Group, there are multiple dust lanes
in addition to stellar shells and tidal debris \citep[][and references
  therein]{2004A&A...423..965B, 2012ApJ...753...43K, 2018ApJ...864..149S}.


We noticed that all five of the clearly merger-built galaxies in our sample 
(NGC~1194, NGC~1316, NGC~2960, NGC~5018, and NGC~5128) reside 
close to each other in the 
$M_{\rm bh}$-$M_{\rm e,sph,eq}$ diagram and the 
$M_{\rm bh}$-$M_{\rm *,sph}$ diagram, where they sit at the 
low-mass end of the elliptical galaxy sequence (Fig.~\ref{Fig-M-R}). 
They have spheroid masses larger than the other
(less gas-rich) S0 galaxies hosting a similar mass black hole.  While not on
the $M_{\rm bh}$-$R_{\rm e,sph}$ relation for elliptical galaxies, they have 
larger sizes than the spheroidal component of the other (relatively gas-poor)
S0 galaxies hosting a similar mass black hole (Fig.~\ref{Fig-M-R}, left-hand
panel).

The red coloured track in the middle of 
Fig.~\ref{Fig-M-R} shows how an equal-mass spiral$+$spiral galaxy merger could 
explain the location and movement of the above five bulges in the
$M_{\rm bh}$-$M_{\rm *,sph}$ diagram.  The associated shift in the 
$M_{\rm bh}$-$R_{\rm e,sph}$ diagram stems from the expected movement along the tight
$M_{\rm *,sph}$-$R_{\rm e,sph}$ relation \citep[][their
  Fig.~7]{Graham:Sahu:22a}.  While the black hole mass in this example
merger (red track) 
has roughly doubled, the spheroid mass has increased four-fold. 
The spheroid mass can more than double depending on how
much of the pre-existing disc stars contribute to the new galaxy's bulge.
The current example could represent the merger of two spiral galaxies with 
bulge-to-total, $B/T$, stellar mass ratios
equal to 1 to create an S0 with $B/T=1/4$.  That is, half of the stars in the
new bulge came from what was disc material, and the other half came from what
was pre-existing bulge material.  Of course, other scenarios are possible,
such as a collision involving an S and an S0 galaxy with different $B/T$
ratios and fractions of discs stars making it into the merger product's bulge. 

Wet mergers are expected to be accompanied by star formation and AGN fueling
\citep[e.g.,][]{1985MNRAS.214...87J,2007A&A...468...61D}, which the `quasar
mode' of galaxy/(black hole) coevolution tries to capture
\citep{2000MNRAS.311..576K, 2005Natur.433..604D, 2006MNRAS.365...11C,
  2013ApJ...772..112S}.  In Fig.~\ref{Fig-M-R}, we only show the jump
(punctuated equilibrium) in growth from the (dissipationless component of the)
merger and not the movement arising from star formation and AGN feeding.
However, `quasar mode' activity may ultimately progress the systems further
along the $M_{\rm bh}$-$M_{\rm *,sph}$ distribution for either S galaxies or
perhaps the relatively (cold gas)-poor S0 galaxies.  Here, we recast this term
(quasar mode) to encapsulate black hole growth from (i) the subsequent binary
black hole merger and (ii) accretion --- as per the original definition ---
and for spheroid growth from (i) the merger of the pre-existing spheroids,
(ii) the fraction of pre-existing disc stars converted into the new spheroid,
and (iii) the fraction of cold gas that underwent star formation to create new
spheroid stars.  Collectively, this is more complicated than the simple
prescriptions given to date \citep[e.g.,][their
  Equation~6]{2013ApJ...764..151G}.  One also needs to resolve where
these wet mergers ultimately end up, that is, do they remain on the E sequence
or evolve to the S or S0 sequence depicted in Fig.~\ref{Fig-M-R}.

The above type of merger event introduces a deviation to the general picture 
presented in \citet{Graham:Sahu:22a}, in which, at a given 
spheroid stellar mass, the bulges of S0
galaxies have an $M_{\rm bh}/M_{\rm *,sph}$ ratio that is an order of
magnitude greater than that in E galaxies.  We now have a situation where, at a
given spheroid stellar mass, the relatively (cold gas)-poor S0 
galaxies have an $M_{\rm bh}/M_{\rm *,sph}$ ratio roughly 
an order of magnitude greater than that in S0 galaxies built by wet mergers. 
The presence/absence of hot gas versus cool gas and dust in these S0 galaxies 
helps discriminate their origin. 
For example, the core-S\'ersic S0 NGC~5813 is immersed in a hot gas halo
\citep{2015ApJ...805..112R} that keeps star formation at bay, 
while the S\'ersic S0 NGC~5128 is rich in cool gas and dust 
\citep[e.g.,][]{1993MNRAS.260..844H}.  The latter is clearly formed from a
gas-rich merger, and the partially-depleted core in the former is thought to
suggest a relatively dry merger origin. 

Like the merger-built elliptical galaxies, these
(wet merger)-built bulges roughly follow the same $M_{\rm bh}$-$M_{\rm *,sph}$
relation as the elliptical galaxies.  
However, as noted above, we have not yet incorporated any black
hole or bulge growth due to gaseous processes.  Conceivably, the AGN may be about to 
experience a Seyfert/quasar growth spurt and return the system to the $M_{\rm
  bh}$-$M_{\rm *,sph}$ relation traced by the relatively (cold gas)-poor S0 galaxies, in
accord with trends seen in some simulations 
\citep[e.g.,][]{2017MNRAS.472L.109A, 2017MNRAS.465...32B, 2018MNRAS.481.3118M,
  2022MNRAS.511.5756T}. 
As with tidal interactions which disturb the H{\footnotesize I} gas reservoir
and ignite AGN \citep{2008ApJ...679.1047K}, minor and major mergers can also
liberate H{\footnotesize I} and H{\footnotesize II} gas from their Keplerian
merry-go-round \citep{1989Natur.340..687H, 1995ApJ...448...41H}.  For an S
galaxy with $M_{\rm bh}/M_{\rm *,sph} \sim 10^{-3}$ and $M_{\rm bh}/M_{\rm
  *,gal} \sim 10^{-4}$ \citep[e.g.,][]{Graham:Sahu:22a}, it does not take an
unrealistic amount of cold gas to boost $M_{\rm bh}$ and thus  
$M_{\rm bh}/M_{\rm *,sph}$.  For example, if gas amounting to 0.1 per cent of
the galaxy's stellar mass were accreted onto the black hole, it would boost
$M_{\rm bh}$ by an order of magnitude.

\citet[][their Fig.~1 and references therein]{2022MNRAS.511.5756T} present a
scenario which may reveal the fate of the S0 galaxies built from wet mergers.
Having exceeded a critical (galaxy stellar) mass threshold, where supernova
also becomes less effective at clearing away potential fuel for the AGN, the
mergers may experience a rapid burst of AGN accretion capable of growing the
black hole by a couple of orders of magnitude.  Even if star-formation were to
double the spheroid's stellar mass, this AGN spurt would more than compensate
for the observed disparity between the (cold gas)-rich and the relatively
(cold gas)-poor S0 galaxies in the $M_{\rm bh}$-$M_{\rm *,sph}$ diagram.
Although, as alluded to by \citet{2000MNRAS.311..576K}, in an evolving
Universe in which galaxies are less (cold gas)-rich, the uptick in the $M_{\rm
  bh}$-$M_{\rm *,sph}$ diagram today may not be as pronounced as it perhaps
was in the younger Universe.  As lower-mass spiral galaxies dry up over time, perhaps due to
ram-pressure stripping from their environment or strong supernova feedback
which inhibits the growth of the spheroid, they too may evolve along a more
upward trajectory in the $M_{\rm bh}$-$M_{\rm *,sph}$ diagram, bringing them
onto the distribution seen for the relatively (cold gas)-poor S0 galaxies.

While the above wet mergers reveal that not all S0 galaxies are faded spiral
galaxies, some may be if (cold gas)-rich disc galaxies evolve (and fade) in a
primarily upward direction in the $M_{\rm bh}$-$M_{\rm *,sph}$ diagram.  Of
course, this would also mean that some S0 galaxies may be faded, merger-built 
S0 galaxies.  This could mesh with the observation that while massive S0
galaxies consist of old stars --- and as we will see in the following section,
some of these have likely been built by dry mergers --- the less massive S0
galaxies can have a broader range of stellar ages \citep{2013MNRAS.432..430B,
  2017MNRAS.471.2687B, 2022MNRAS.513..389R}.

\begin{figure*}
\begin{center}
\includegraphics[trim=0.0cm 0cm 0.0cm 0cm, width=1.0\textwidth, angle=0]{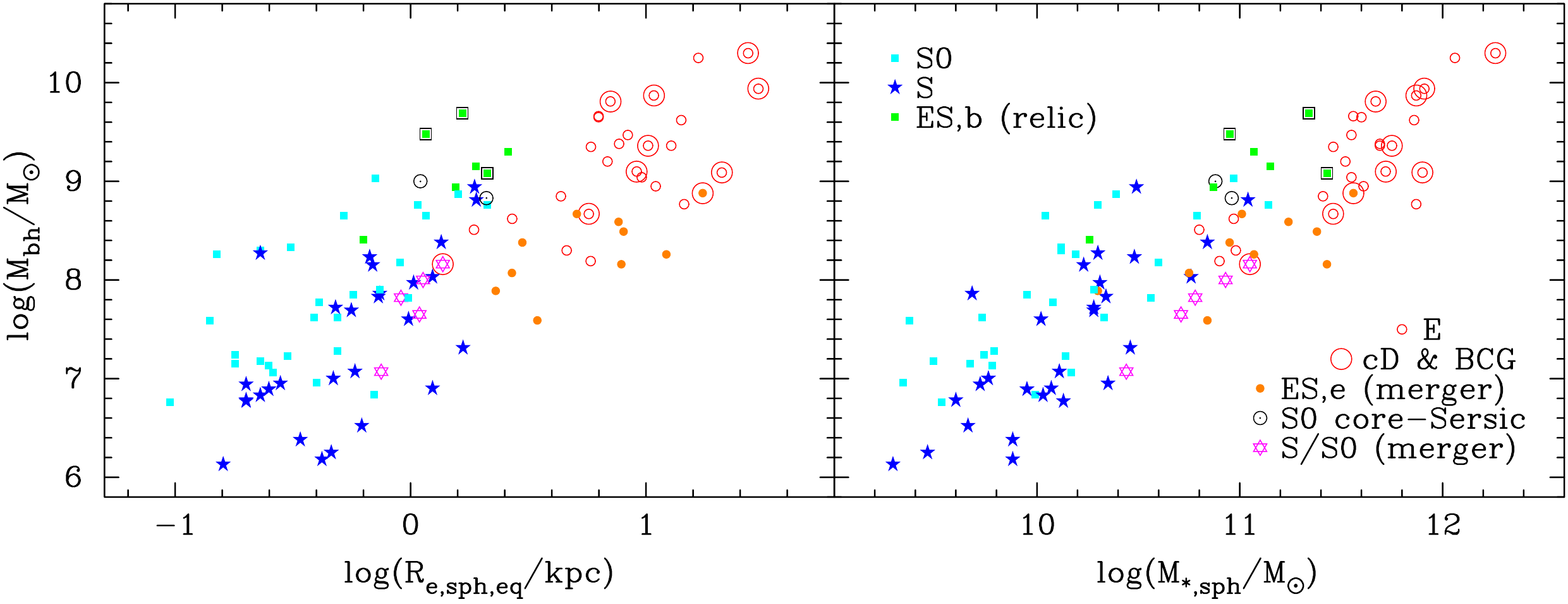}
\caption{Similar to Fig.~\ref{Fig-M-R} but now coded according to a finer 
  division of morphological type, slightly updated for some ETGs as per Table~\ref{TableESeESb}.
Spiral (S) = blue star; 
lenticular from S and S0 merger (S0,pec.) = pink hexagon star; 
S\'ersic lenticular (S0) = cyan square; 
core-S\'ersic lenticular from S0 merger (S0) = dotted black circle; 
relic ellicular (ES,b) = green square; 
merger-built ellicular (ES,e) = orange dot; 
merger-built elliptical (E) = small red circle; 
brightest group/cluster galaxy (BGG/BCG) = large red circle. 
Building on \citet[][their Fig.~5]{2016MNRAS.457..320S}, 
the non-Spitzer ES,b galaxies Mrk~1216, 
NGC~1277 (the ES,b galaxy with the highest spheroid mass, based on $M_*/L_V=11.65$)
and NGC~1271 (Section~\ref{Sec_dry_merger}) 
have been added and enclosed in a black square; 
their masses are taken from \citet[][and references therein]{2020ApJ...903...97S}. 
The (non-BCG) E galaxy offset to the lower-right of the ensemble
of E galaxies in the right-hand panel is the 
Seyfert galaxy NGC~6251 ($\log(M_{*,sph}/M_\odot)=11.87$, 
$R_{\rm e,sph,eq}=14.5$~kpc).   The S galaxy offset to the left in the
left-hand panel is NGC~4699, remodelled in the Appendix and potentially
serving as a warning to the question `what is a bulge?'.
}
\label{Fig-M-R-2}
\end{center}
\end{figure*}


\begin{figure}
\begin{center}
\includegraphics[trim=0.0cm 0cm 0.0cm 0cm, width=1.0\columnwidth, angle=0]{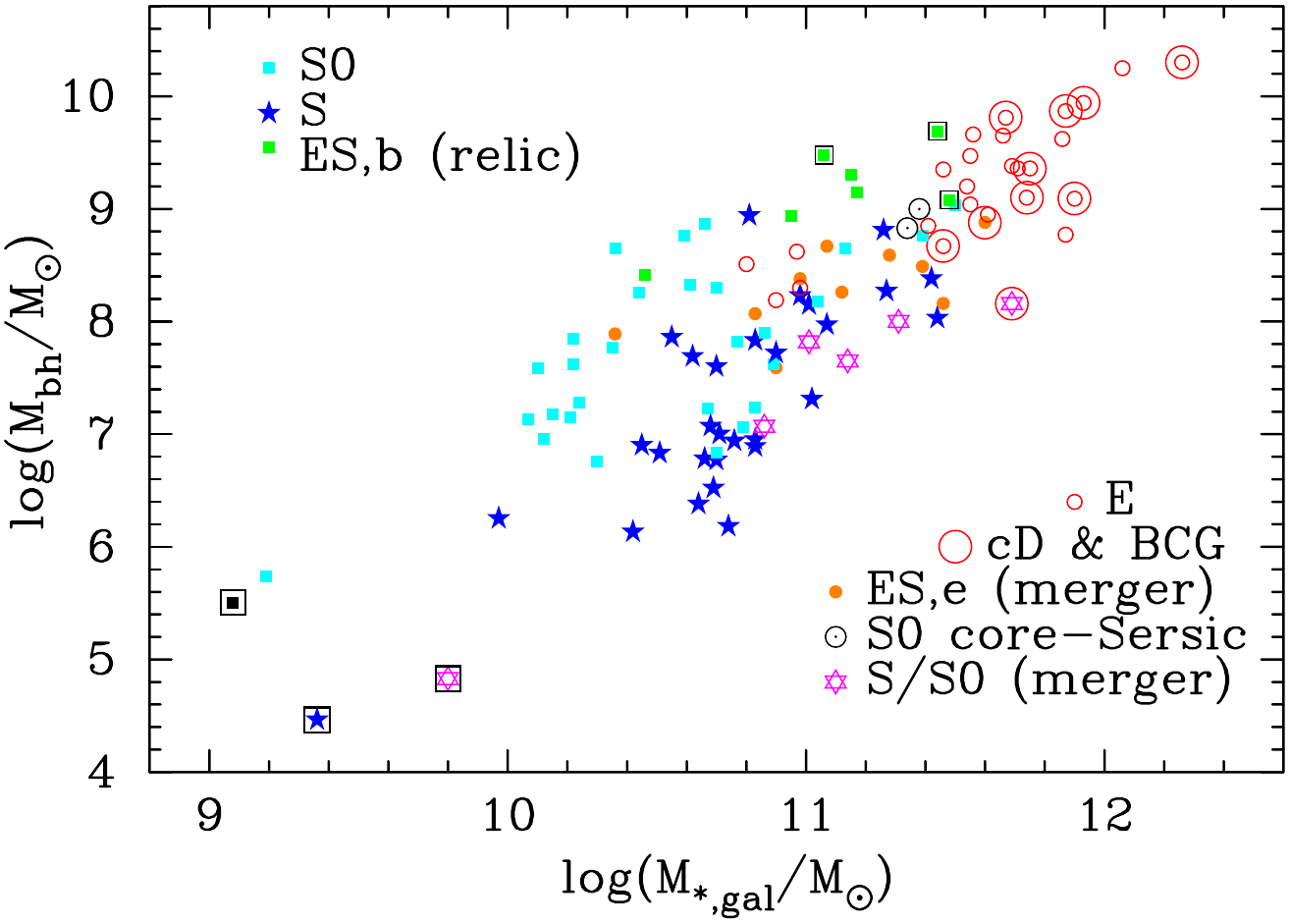}
\caption{Similar to the right-hand panel of Fig.~\ref{Fig-M-R-2} but
  displaying the galaxies' stellar mass.  The galaxies in the lower left are:
  NGC~404 (S0); Pox~52 \citep[ETG:][]{2008ApJ...686..892T}; LEDA~87300
  \citep[S:][]{2015ApJ...809L..14B, 2016ApJ...818..172G}; 
and NGC~4424 \citep[S/merger:][]{2021ApJ...923..146G}. Systems with a
black square around them are not a part of our Spitzer sample and have had
their masses taken from the cited literature.  The ES,e galaxy with the lowest
stellar mass is NGC~3377. The S galaxy with the highest black hole mass is 
NGC~5055; however, this is an error and is actually the mass within the inner
300~pc \citep{2004A&A...420..147B}. 
The S0/merger with the highest stellar mass is NGC~1316 (Fornax~A). 
}
\label{Fig-M-R-3}
\end{center}
\end{figure}

\begin{figure}
\begin{center}
\includegraphics[trim=0.0cm 0cm 0.0cm 0cm, width=1.0\columnwidth, angle=0]{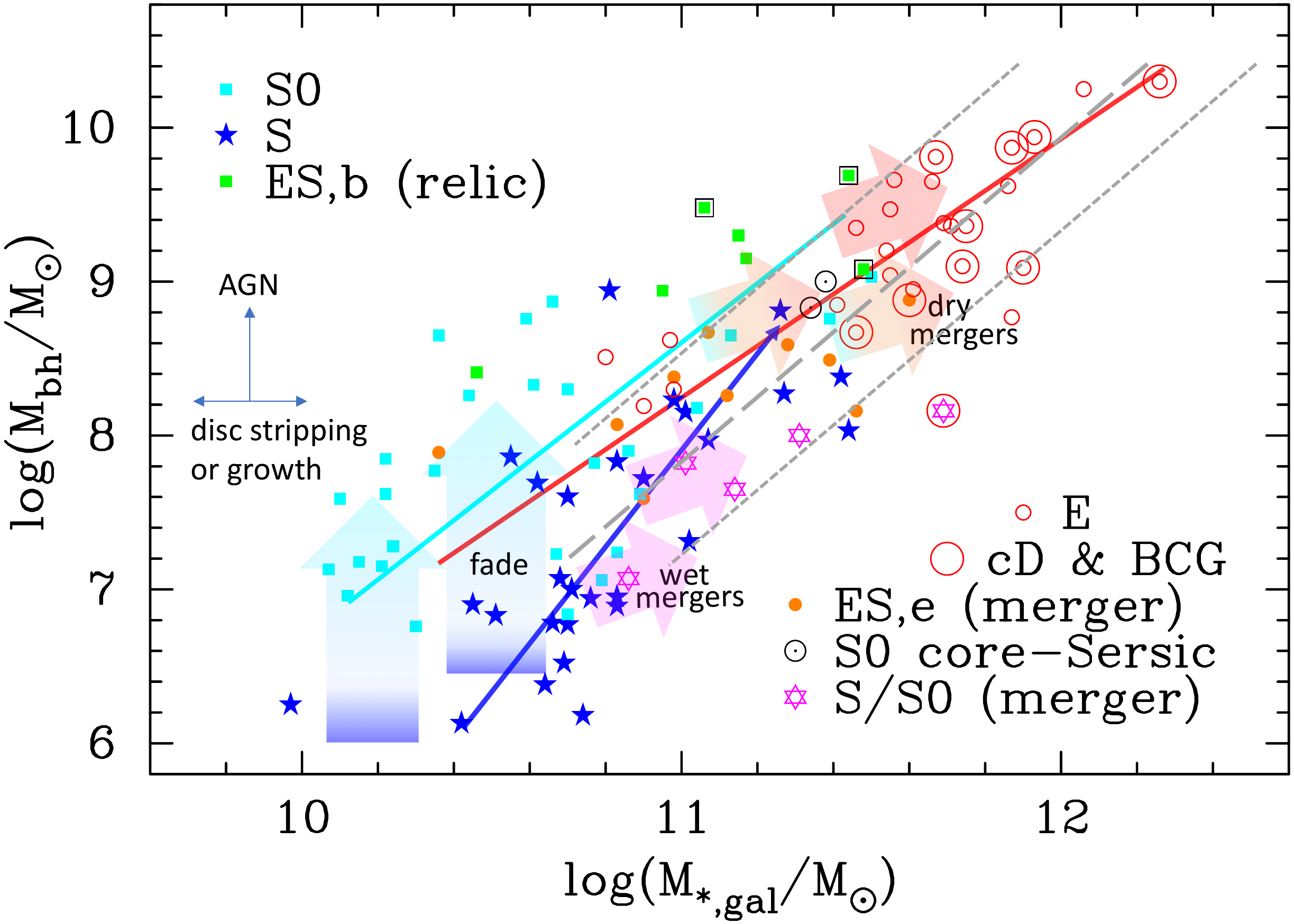}
\caption{Similar to Fig.~\ref{Fig-M-R-3} but slightly zoomed in and showing
  some evolutionary tracks.  The cyan, red and blue lines show the
  relations from \citet[][their Table~2]{Graham:Sahu:22a} for the (cold
  gas)-poor S0, E$+$BCG and S galaxies.  The solid grey line depicts
  Equation~\ref{Eq_merge}, with the dashed grey lines bounding the general
  distribution of merger-built systems.  The broad blue arrows show how some S
  galaxies might evolve by using up their fuel and fading into relatively
  (cold gas)-poor S0 galaxies, which may, in turn, collide to become (cold
  gas)-poor elliptical galaxies (broad red arrows), or the disc galaxies may
  undergo a merger event while still (cold gas)-rich to become dusty/(cold
  gas)-rich S0 galaxies (broad pink arrows).  Rather than a major merger
  event, some S galaxies may continue to accrete material and evolve along the
  S galaxy sequence (dark blue line).  }
\label{Fig-M-R-4}
\end{center}
\end{figure}

We note that the $M_{\rm bh}$-$M_{\rm *}$ diagram from the SIMBA
simulation \citep[][their Fig.~13]{2019MNRAS.486.2827D}, when coloured
according to the (galaxy) specific star formation rate, shows a promising
correlation with what is expected given the galaxy types shown in
Figures~\ref{Fig-M-R-2} and \ref{Fig-M-R-3}.  This agreement, at least at
$M_{\rm bh}/M_\odot \gtrsim 10^7$, is even more evident in Fig.~4 from
\citet{2021MNRAS.503.1940H}.  The merger-induced jump, explained in
\citet{Graham:Sahu:22a}, from the non-(star-forming) S0 galaxies defining the
left-hand envelope of points across to the E galaxies is even more pronounced
in the $M_{\rm bh}$-$M_{\rm e,sph}$ diagram (Fig.~\ref{Fig-M-R-2},
left-hand side). 

If not for the ES,b and S0 galaxies (the green and cyan squares in
Fig.~\ref{Fig-M-R-3}), there would be a notably tighter $M_{\rm bh}$-$M_{\rm
  *,gal}$ relation.\footnote{Removal of the ES,b and S0 galaxies also reduces
  the scatter in the $M_{\rm bh}$-$M_{\rm *,sph}$ and $M_{\rm bh}$-$M_{\rm
    e,sph}$ diagrams (Figure~\ref{Fig-M-R-2}.} 
  If the ES,b galaxies were to acquire a
large-scale disc and thus have a higher disc-to-bulge stellar mass ratio
more typical of S0 galaxies, 
they would reside closer to the E galaxies. This makes us wonder
if some of the S0 galaxies, specifically the cyan points on the left-hand side
of the distribution in Fig.~\ref{Fig-M-R-3}, may have reduced disc-to-bulge ratios.
Might they perhaps reside in clusters, where S0 galaxy disc scalelengths are
known to be smaller \citep[e.g.,][]{2004ApJ...602..664G}, perhaps due to
cluster-hindered growth from ram pressure stripping of gas 
\citep[e.g.,][]{1972ApJ...176....1G, 2010AJ....140.1814Y, 2013MNRAS.429.1747M}
or tidal stripping of disc stars \citep{1996Natur.379..613M} which build the
intracluster light (ICL). 
The BCGs aside, the high galaxy velocities in clusters will act to hinder collisions,
thereby preventing some systems from moving across to the right in the 
$M_{\rm bh}$-$M_{\rm *,gal}$ diagram.  Supernovae and stellar
winds \citep[e.g.,][]{1977A&A....61..251D,2022arXiv220308218B} may also reduce the stellar mass, 
while a final burst of quasar fuelling might drive the systems up in the $M_{\rm bh}$-$M_{\rm
  *,gal}$ diagram before they `dry out'.  In future work, the S0 galaxies 
will be studied in more detail, as will the addition of 
intermediate mass black hole (IMBH) candidates 
\citep[e.g.,][]{2018ApJ...863....1C, 2019MNRAS.484..814G,
  2020ApJ...898L..30M}.



With an eye to collisions involving supermassive black holes,
Fig.~\ref{Fig-M-R-4} may offer some guidance for simulations and
predictions for the stochastic background of long-wavelength gravitational
radiation.  
Due to the clutter in Fig.~\ref{Fig-M-R-4}, it was felt that a less
adulterated version (Fig.~\ref{Fig-M-R-3}) should be included. 
Although the number of data points is still low, it appears that
the merger-built lenticular galaxies broadly partake in the
same distribution as the elliptical galaxies.\footnote{Curiously, one can also
  see how removing the cyan points, i.e., the possibly (non-merger)-built S0
  galaxies, will result in a notably tighter $M_{\rm 
    bh}$-$M_{\rm *,gal}$ relation.  This will be pursued in a future paper.}  
This observation may be of practical use if (i) all
(major merger)-built galaxies in the $M_{\rm bh}$-$M_{\rm *,gal}$ diagram
follow the same relation and (ii) no bulge/disc/etc.\ decomposition is
required.  To date, obvious mergers have been excluded from the scaling
relations, and it has frequently been emphasised that black holes correlate with spheroid
mass rather than galaxy mass (e.g., KH13).  However, a 
merger-built galaxy relation represents the `end product' of major mergers
generating nanohertz gravitational waves \citep[e.g.,][]{2019A&ARv..27....5B}. 
Fig.~\ref{Fig-M-R-4} may reveal how elliptical galaxies are built from dry
mergers involving lenticular galaxies on the (cold gas)-poor S0
`sequence'\footnote{This `sequence' is perhaps more of an elongated cloud until lower
  mass S0 galaxies are included.} while 
relatively (cold gas)-rich S0,pec galaxies are built from wet mergers.
This broad merger-built galaxy relation has not been presented before and has
a steeper slope than 1.69$\pm$0.17 defined by just the E galaxies in 
\citet{Graham:Sahu:22a}. It is instead roughly such that 
\begin{equation}
\log \frac{M_{\rm bh}}{M_\odot} = 2.2\log \frac{M_{\rm *,gal}}
 {\upsilon \, 10^{11} \, M_\odot}+7.8, 
\label{Eq_merge} 
\end{equation}
with $\upsilon=1$ when using stellar mass-to-light ratios $\Upsilon_*$ that are consistent
with \citet[][their Equation~4]{Graham:Sahu:22a}, which is based on the models of 
\citet{2013MNRAS.430.2715I} and a \citet{2002Sci...295...82K} initial mass
function.  Equation~\ref{Eq_merge} is shown by the grey line in
Fig.~\ref{Fig-M-R-4}.

\subsubsection{Alleged pseudobulges}\label{Sec_pseudo}

\citet{2001ASPC..230..247K} 
initially reported that classical and pseudobulges follow the same relation. 
\citet{2007AAS...211.1327G} and \citet{2008ApJ...680..143G} 
observed that barred galaxies, often associated with
pseudobulges \citep[e.g.,][]{2011MNRAS.415.3308G}, 
appeared offset from the relation defined by the non-barred galaxies
in the $M_{\rm bh}$-$\sigma$ diagram \citep{2000ApJ...539L...9F,
  2000ApJ...539L..13G}. 
This was the first paper to report on potential 
morphology-dependent substructure in the black hole scaling diagrams. 
\citet{2008MNRAS.386.2242H} 
reported a morphology-dependence in terms of classical bulges versus 
pseudobulges; although all of their alleged pseudobulges were barred galaxies, 
\citep[see also ][]{2009MNRAS.399..621G}.
\citet{2011Natur.469..374K} subsequently 
embraced this notion of an offset population of pseudobulges with low black
hole masses relative to their bulge mass, while 
\citet{2011MNRAS.412.2211G} continued to favour an offset population of barred galaxies
with elevated velocity dispersions --- a reported feature of barred galaxies
\citep{2005ApJ...629..797G, 2014MNRAS.441.1243H, 2013ApJ...778..151B, 2013ApJ...765...23D}. 

Complicating matters was that the problematic identification \citep{2014ASPC..480..185G} of
pseudobulges is a subjective practice leading to classification disagreements
in the literature, as noted in, for example, 
\citet[][their Section~4.1]{2017MNRAS.471.2187D}.
%
%
Furthermore, some galaxies (e.g., NGC~2787, 
NGC 3368, NGC 4371 and NGC 4699) are claimed to have both types of bulge
\citep[e.g.,][]{2003ApJ...597..929E, 2015MNRAS.446.4039E,
  2015A&A...584A..90G}, and simulations have shown how this can occur
\citep[e.g.,][]{2016ApJ...821...90A}. 
Our use of multicomponent decompositions circumvents many instances of
pseudobulge confusion because the alleged pseudobulge, whether an inner
disc or (peanut shell)-shaped structure associated with the bar, 
is captured, modelled, and not considered the bulge.  For example, 
as seen in Appendix~\ref{Apdx1}, 
we remodel NGC~2787 by introducing a truncated-exponential for the inner
disc, i.e., the purported pseudobulge in this galaxy \citep{2003ApJ...597..929E}.
We also remodelled NGC~4699, whose previous `pseudobulge' was a conglomeration of
the bar, spiral arms, and disc \citep{2015MNRAS.446.4039E}, none of which we
consider to be the `bulge' component. 
In addition, as noted earlier, boxy/X/(peanut shell)-shaped perturbations
arising from bar instabilities, and labelled as pseudobulges by some,
are not treated as bulges in our image analysis (except, of course, for those we
have undoubtedly, inadvertently, over-looked).

Over time, as the sample size increased, the picture cleared, and
\citet{2019ApJ...887...10S} revealed that not all barred galaxies are offset
to higher velocity dispersions.  All bulges appear to follow the same
$M_{\rm bh}$-$\sigma$ relation, at least until higher masses where the
relation steepens \citep{2018ApJ...852..131B}, 
perhaps associated with the onset of partially depleted
cores \citep{2019ApJ...887...10S} or for virialisation issues that will be
dealt with in a subsequent paper.  Now, neither the barred galaxies nor the
alleged pseudobulges appear to form an offset population with low $M_{\rm bh}$
in the $M_{\rm bh}$-$\sigma$ diagram.

KH13 (their Table~3) excluded roughly half (22/43) of the  bulges in their parent
sample, deeming them `pseudobulges' rather than `classical' bulges. 
In passing, we clarify that KH13 excluded an additional four `classical'
bulges (NGC~1277, NGC~3998, NGC~4342 and NGC~4596), therefore explaining why
their bulge sample used for their linear regression was reduced to 17 rather
than 21, i.e., they used 40 per cent of the parent sample of 43 bulges. 
Of these four classical bulges, they rightly avoided using NGC~1277 at that
time. The $M_{\rm bh}/M_{\rm *,sph}$ ratio in their table was later shown to
be too high by two orders of 
magnitude \citep{2016ApJ...819...43G}, and the system is no longer considered unusual. 
This galaxy was not a part of our Spitzer sample in
\citet{2019ApJ...876..155S}; however, we plot it in
Fig.~\ref{Fig-M-R-2}, where it {\em appears} that the $M_*/L_V$ ratio
used to obtain the spheroid mass may be too high.  This ratio is
revisited in the following subsection. 
Finally, NGC~4342 is a stripped galaxy \citep{2014MNRAS.439.2420B} that we also exclude
from the Bayesian linear regression analysis in \citet{Graham:Sahu:22a}
but plot in Fig.~\ref{Fig-M-R}.  
We do, however, include NGC~3998 \citep{2006A&A...460..439D, 2018MNRAS.473.2930D}
and NGC~4596 (see Section~\ref{Sec_BCGs}).

In effect, KH13 (see their Fig.~21) exclude all but four systems with black
hole masses less than $\sim$$3\times10^7$ M$_\odot$ (plus four alleged
pseudobulges with $M_{\rm bh}$$\approx$(3--8)$\times10^7$ M$_\odot$).  Not
surprisingly, with their select pseudobulge sample spanning just 1.2 dex
(NGC~4395 aside), they reported the absence of any correlations among that sample.  However, as
revealed in, for example, Fig.~\ref{Fig-M-R-2}, these low-mass systems are 
involved in strong correlations. 


Although we have tried to be inclusive, 
we, too, have a sample selection at play, such that spiral galaxies with apparently
small bulges are excluded.  Specifically, 13 spiral galaxies whose bulge required
a spatial resolution better than the Spitzer Space Telescope could provide
have not been included because their galaxy analyses was performed on Hubble
Space Telescope F814W images, for which we may not have a consistent
prescription for $\Upsilon_*$.  
However, it is thought that this exclusion of the, in
general, more distant spiral galaxies is not introducing a bias, given that the
near-quadratic $M_{\rm bh}$-$M_{\rm *,sph}$ relation obtained using the
sample of spiral galaxy bulges with Spitzer data \citep{Graham:Sahu:22a} is consistent
with that obtained from the fuller sample of spiral galaxy bulges containing
the above exclusions \citet{2019ApJ...873...85D}.
Furthermore, it is noted in passing that galaxies with a direct black hole
mass measurement do not represent a biased sample, in the (galaxy stellar mass)
--(central velocity dispersion) diagram, with respect to those not
having a direct measurement \citep{SahuGrahamHon22}.  In addition, the
overlapping agreement seen in the spheroid size-mass diagram for samples with
and without a direct black hole mass measurement \citep{HGS2022} also reveals
no sample selection bias.  As such, neither the $M_{\rm bh}$-$M_{\rm *,sph}$, 
$M_{\rm bh}$-$R_{\rm e,sph}$, 
nor $M_{\rm bh}$-$\sigma$ relations are thought to be biased. 
However, both bulge- and disc-dominated galaxies with low surface brightness discs 
\citep[e.g.,][]{2001ApJ...556..177G} 
but galaxy stellar masses comparable to those seen in 
Figures~\ref{Fig-M-R-3} and \ref{Fig-M-R-4} are absent.


\subsection{Spheroids at the top of the bulge sequence}
\label{Sec_dry_merger}

As discussed in \citet{Graham:Sahu:22a}, massive compact spheroids from `relic
galaxies'  --- largely unevolved, compact massive galaxies, aka `red nuggets' from
$z\sim2.0\pm0.5$ with $M_* \gtrsim 10^{11}$ M$_\odot$, $R_{\rm e} \lesssim
2$~kpc, 
and a red colour characteristic of a quiescent non-starforming stellar
population \citep[][and references therein]{2005ApJ...626..680D,
  2009ApJ...695..101D, 2022MNRAS.514.3410H} --- tend to reside near the
top of the bulge sequence in the $M_{\rm bh}$-$M_{\rm *,sph}$ diagram. 
Examples are the ES,b galaxies NGC~1332 and NGC~5845 listed in
Table~\ref{TableESeESb}. 
%
%
Other examples, albeit without our uniform analysis of a Spitzer image, 
include 
Mrk~1216 
\citep[$\log(M_{\rm bh}/M_{\odot}) \le 9.69\pm0.16$, D=40.7~Mpc,][]{2015MNRAS.452.1792Y} 
with 
\citep[$\log(M_{\rm *,sph}/M_{\odot})=11.34\pm0.20$,][]{2016MNRAS.457..320S}, 
NGC~1271 
\citep[$\log(M_{\rm bh}/M_\odot)=9.48\pm0.16$,][]{2015ApJ...808..183W}
with 
\citep[$\log(M_{\rm *,sph}/M_\odot)=10.95\pm0.10$,][]{2016ApJ...831..132G}, 
and NGC~1277 
with $\log(M_{\rm *,sph}/M_\odot)=11.43\pm0.10$
\citep[central $M_*/L_V=11.65$,][]{2015MNRAS.451.1081M} 
and a likely upper limit 
to the black hole mass of $1.2\times10^9$ M$_\odot$ based on high spatial
resolution  
(adaptive optics)-assisted Keck/{\sc osiris} data \citep{2016ApJ...819...43G}. 
Looking at Fig.~\ref{Fig-M-R-2}, the spheroid mass for NGC~1277 seems
some three times higher than expected from the distribution of the other ES,b galaxies.
Although 
\citet[][their Fig.~6]{2015MNRAS.451.1081M} suggest a lower $M_*/L_V \approx 8$
from 0.7 to 1.4 $R_{\rm e,gal}$, this would only reduce the spheroid mass by one-third.
It would require $M_*/L_V \approx 4$ to yield a two-third reduction giving
the speculated factor of three shift.

NGC~5252 
\citep[$\log(M_{\rm bh}/M_\odot)=9.03\pm0.40$,][]{2005A&A...431..465C}, 
with 
$\log(M_{\rm *,sph}/M_\odot)=10.97\pm0.27$ \citep{2019ApJ...876..155S}, 
is thought to represent a (possible relic) compact massive spheroid that has
acquired a large-scale disc rather than an intermediate-scale disc 
\citep[see][and references therein]{2015ApJ...804...32G}.  It 
is the S0 galaxy with the highest black hole mass among the S0 galaxies in our 
sample and is thought to represent 
a good example of the two-step scenario introduced\footnote{First posted on
  arxiv.org in 2011.} by 
\citet{2013pss6.book...91G} in which discs grow in and around pre-existing
`red nuggets'.  The potential binary black hole in this galaxy would be
further evidence of a past accretion event \citep{2017MNRAS.464L..70Y}. 

NGC~6861 is an ES,b galaxy likely to cause some confusion, assuming we have
assigned the correct designation.  While it appears to have grown/accreted an 
intermediate-scale disc \citep{2019ApJ...876..155S}, there is still a dusty
disc in this galaxy, at odds with what the term `relic' might conjure for some.
It is. however, a galaxy that may have been left behind in an evolutionary
sense, having not acquired a substantial disc mass. 



\subsubsection{Bulges with core-S\'ersic profiles}

Dry galaxy mergers building `fast rotators', specifically S0 disc galaxies
\citep[e.g.,][]{2013MNRAS.432.1768K}, are readily explained in simulations
\citep{2014MNRAS.444.3357N}.  The S0 galaxies with partially depleted cores
have likely been built by inspiralling black holes in relatively dry mergers
\citep{1991Natur.354..212E}, whereas relic red nuggets are not expected to
have depleted cores, nor are S0 galaxies built from wet mergers, like
NGC~5128, which may yet evolve upward in the $M_{\rm bh}$-$M_{\rm sph}$
diagram to eventually join the (cold gas)-poor S0 galaxies, or perhaps the
(cold gas)-poor S0 galaxies have been frozen in time, reflective of how things
scaled long ago.  

The core-S\'ersic S0 galaxies in our sample do not appear to reside on the
elliptical galaxy sequence in the $M_{\rm bh}$-$M_{\rm sph}$ diagram.  This
is expected when the merger event does not fold in much of the progenitors'
disc mass.  It would not, however, be unreasonable to expect (as more data
becomes available) a continuum --- based on the amount of disc mass converted
into bulge mass --- of relatively dry merger remnants bridging the left- and right-hand
sides of the distribution seen in Fig.~\ref{Fig-M-R-2}.

Five core-S\'ersic S0s are identified in \citet{2019ApJ...887...10S}.  They
are NGC~524, NGC~584, NGC~3796, NGC~4751 and NGC~5813, with the first and last
galaxies appearing in our Spitzer sample.  These two have been identified in
Fig.~\ref{Fig-M-R-2}.  Their location can, once again, be explained by a
merger in which the net orbital angular momentum of the pre-merged progenitor
galaxies --- themselves likely to be lesser lenticular galaxies --- is not
cancelled.  It is worth reiterating that these core-S\'ersic galaxies with
large-scale discs are not thought to have formed from an overly gas-rich
dissipative merger event, otherwise the depleted stellar core may not have
formed.  This would imply that the `quasar mode' of galaxy/AGN growth did not
dominate the construction of the spheroid in these disc galaxies.  Although,
dust is evident\footnote{See the `level 4' colour images available from the Hubble
  Legacy Archive (HLA: \url{https://hla.stsci.edu}).} 
in NGC~524, with its dusty spiral arms/rings, and in NGC~5813,
and thus there was some gas and past star formation.  Furthermore, at least in
NGC~5813, nuclear dust may complicate the identification of the depleted core,
an issue highlighted in two other galaxies by \citep{2018MNRAS.478.1161B}. 
%

Galaxies without cold gas, especially those of high-mass embedded within a hot
gas halo \citep[e.g.,]{1994Natur.369..462P, 2005MNRAS.364..169P}, would be
expected to evolve further only through mergers.  The `hot mode' / `radio
mode' activity maintained by the central `Benson burner' \citep[a term
  introduced by][see their footnote~33]{Graham:Sahu:22a} in these galaxies
performs a maintenance, rather than an evolutionary, role that keeps a 
galaxy `dry', by which we mean essentially free of cold gas capable of forming
stars.  However, significant `cold stream' inflows from the intergalactic
medium could infiltrate the galaxy and begin to build a new or grow an
existing disc \citep{2005MNRAS.363....2K, 2009Natur.457..451D,
  2013Sci...341...50B}.

One significant difference between the merger of low- and high-mass disc
galaxies is that the pre-merged low-mass galaxies will have smaller $B/T$
ratios, and thus, there is greater scope for larger fractional increases in
the bulge mass of the post-merger end-product.  This is reflected by the
longer horizontal arrow of the middle red track in Fig.~\ref{Fig-M-R}
relative to the upper red track in this figure.

\begin{table*}
\centering
\caption{ES galaxies}\label{TableESeESb}
\begin{tabular}{llcllrcll}
\hline
Galaxy    &  Type & $\log(M_{\rm bh}/M_\odot)$ & $M_{\rm bh}/M_{\rm *,sph}$ & $\rho_{\rm e,sph}$  & $R_{\rm e,sph,eq}$ &  $n_{\rm sph,eq}$ & Ref & Notes \\
          &       &                 &       & $M_\odot/{\rm pc}^3$ & kpc &  &  &  \\ 
\hline 
NGC~0821  &  ES,e &  7.59 & 0.00056 & 0.40  & 3.5  &  5.2 & Fig.~\ref{Fig-N821}   &    \\ 
NGC~1275  &  ES,e &  8.88 & 0.0021  & 0.016 & 17.4 &  4.3 &  SGD19 &  cD, ES,pec.\ with undigested component. \\
NGC~3377  &  ES,e &  7.89 & 0.0039  & 0.39  & 2.3  &  4.5 &  Fig.~\ref{Fig-N3377}  & Offset in some diagrams  \\
NGC~3414  &  ES,e &  8.38 & 0.0027  & 0.80  & 3.0  &  4.5 &  SG16  &  ES,pec.\ with a polar ring possibly building a disc. \\
NGC~3585  &  ES,e &  8.49 & 0.0013  & 0.11  & 8.0  &  6.3 &  SG16  & High-rotation \citep{1995AnA...293...20S}. \\  
NGC~3607  &  ES,e &  8.16 & 0.00054 & 0.13  & 7.9  &  5.6 &  SG16  &  \\
NGC~4473  &  ES,e &  8.07 & 0.0021  & 0.69  & 2.7  &  2.9 &  SG16  &  Two counter-rot.\ inner discs.  \\ 
NGC~4552  &  ES,e &  8.67 & 0.0046  & 0.19  & 5.1  &  5.4 &  SGD19 &  Inner ring and undigested component. \\
NGC~4621  &  ES,e &  8.59 & 0.0022  & 0.09  & 7.6  &  8.8 &  SG16  &  M59. \\
NGC~4697  &  ES,e &  8.26 & 0.0015  & 0.015 & 12.2 &  6.7 &  SG16  &  BGG in NGC~4697 Group. \\
\hline 
NGC~1332  &  ES,b &  9.15 & 0.010   & 4.99  & 1.9  &  3.7 &  SG16  &  Maybe an S0 \citep{2011MNRAS.410.1223R}. \\
NGC~3115  &  ES,b &  8.94 & 0.012   & 4.75  & 1.6  &  5.1 &  SG16  &  \\
NGC~5845  &  ES,b &  8.41 & 0.014   & 17.4  & 0.6  &  3.3 &  SGD19 &  \\
NGC~6861  &  ES,b &  9.30 & 0.017   & 1.59  & 2.6  &  3.5 &  SGD19 &  \\
\hline
\end{tabular}

These tabulated ES galaxy types may reflect the galaxies' history.  They are based on the
$M_{\rm bh}/M_{\rm *,sph}$ ratio, the spheroid's stellar density, $\rho_{\rm
  e,sph}$, within the spheroid's effective half light radii, $R_{\rm
  e,sph,eq}$, and the galaxy appearance (not to be confused with the
spheroid S\'ersic index, $n_{\rm sph,eq}$, shown here for completeness).
The parameters are sourced from \citet{Graham:Sahu:22a}.  ES,e: galaxies may
be akin to merger-built E galaxies but with an inner intermediate-scale disc.
NGC~4594 also belongs to this type.  
ES,b: galaxies can be thought of as bulges which did not grow up: relic `red
nuggets' which did not accrete a large-scale disc.  The values $n_{\rm
  sph,eq}$ and $R_{\rm e,sph,eq}$ pertain to the equivalent-axis,
i.e., the geometric-mean axis, and have the usual meaning 
\citep[e.g.,][]{1993MNRAS.265.1013C, 2005PASA...22..118G}, with half the
spheroid's light contained within $R_{\rm e,sph,eq}$. 
The external references showing our
decompositions of the galaxy light are: SG16 = \citet{2016ApJS..222...10S};
SGD19 = \citet{2019ApJ...876..155S}.

\end{table*}

\subsubsection{Massive ES galaxies}\label{Sec_ESe-b}

There has been a tendency to misidentify ES galaxies in the literature.  They
can be dwarf galaxies \citep{2012ApJ...750..121G, 2017ApJ...840...68G} or
massive galaxies \citep{2016MNRAS.457..320S}.  Their discs are often missed,
and the galaxies are modelled with a single S\'ersic function.  Or, when
the disc is detected, they 
tend to be modelled as large-scale discs, dominating the light
at large radii.  However, ES galaxies have intermediate-scale discs, with the spheroid
dominating the light at large radii \citep{1966ApJ...146...28L,
  2016MNRAS.457..320S}.  Just as the E and S0 classification does
not apply to them, neither do the terms `slow rotator' or `fast rotator'. 
ES galaxies are both fast and slow rotators depending on the radial range
sampled \citep{2017MNRAS.470.1321B}.  Both wet and dry 
mergers can build fast rotators if the net angular momentum is not nulled, and
it stands to reason that both types of mergers can build ES galaxies.  As
described in \citet{Graham:Sahu:22a}, the 
slow rotating E galaxies appear to be built from dry mergers involving ES and S0
galaxies with larger $B/T$ ratios. 

In \citet{Graham:Sahu:22a}, our focus was on the broad
division between bulges and merger-built E galaxies.  Half a dozen ES
galaxies (NGC: 821; 3377; 3607; 4473; 4621; 4697) were modelled as such, i.e.,
as ES galaxies, but treated as E galaxies (with inner discs) for consistency
with the designations in \citet{2019ApJ...876..155S}.  The ES galaxies in our
sample have spheroid-to-total flux ratios from 0.8--0.85 to 0.95.  Here, with
our higher resolution investigation of the $M_{\rm bh}$-$M_{\rm *,sph}$
diagram, in which we attempt to read the proverbial tea leaves and go beyond
the first-order divisions reported by \citet{2019ApJ...876..155S}, we look more
closely at the ES galaxies (listed in Table~\ref{TableESeESb}) and their
possible subdivision.

There appears to be two types of ES galaxies with an intermediate-scale
disc.  One of these are the relic `red nuggets' 
\citep[e.g.,][]{2017MNRAS.467.1929F} 
from the high-$z$ Universe that have
evolved little and not accreted a substantial disc 
(Section~\ref{Sec_dry_merger}), 
designated here as ES,b. They may be expected to 
reside at the top of the bulge sequence in the $M_{\rm bh}$-$M_{\rm *,sph}$
diagram.  Somewhat midway between core-S\'ersic S0 galaxies and disc-less
elliptical galaxies are merger-built ES,e galaxies. They may have transformed
more of their progenitor galaxies' disc stars into a spheroid than was
achieved with the merger-built S0 galaxies, i.e., the core-S\'ersic S0
galaxies, but did not achieve the high conversion level required to become an
E galaxy.   Many of the examples in Table~\ref{TableESeESb} have previously
been identified as ``disc elliptical'' \citep[][their Table~1]{1995AnA...293...20S}.
We summarise this as follows. 

(i) ES,e: merger-built, near-elliptical galaxies with partial disc formation.
These spheroids and galaxies 
 are not expected to be compact ($R_{\rm e} \lesssim 2$ kpc) 
in size like 'red 
nuggets' and can be considered more in the E galaxy camp than the bulge camp.
They will have reduced $M_{\rm bh}/M_{\rm *,sph}$ ratios due to the pre-merged
galaxies' disc stars bolstering the spheroid mass of the merger-built galaxy. 

(ii) ES,b: relic `red nuggets' with partial disc accretion.  These spheroids
should still be compact and can be considered more in the bulge camp than the
E galaxy camp.  They will be the rare compact massive systems largely
unevolved since $z\sim2$, such as NGC~1277 \citep{2014ApJ...780L..20T,
  2016ApJ...819...43G}.  NGC~3115 \citep{1940ApJ....91..273O,
  2011ApJ...736L..26A}, with $R_{\rm e,sph,eq}=1.55$ kpc and $\log(M_{\rm
  *,sph}/M_{\odot})=10.87$ \citep{Graham:Sahu:22a} may be another example.  In
this case, it may be a `red nugget' which experienced a single minor merger
with a small spiral galaxy that created the intermediate-scale disc.  Compared
to the ES,e systems, the ES,b spheroids may have higher stellar densities
within their half-light radii (see Table~\ref{TableESeESb}).


\begin{figure}
\begin{center}
\includegraphics[trim=0.0cm 0cm 0.0cm 0cm, width=1.0\columnwidth, angle=0]{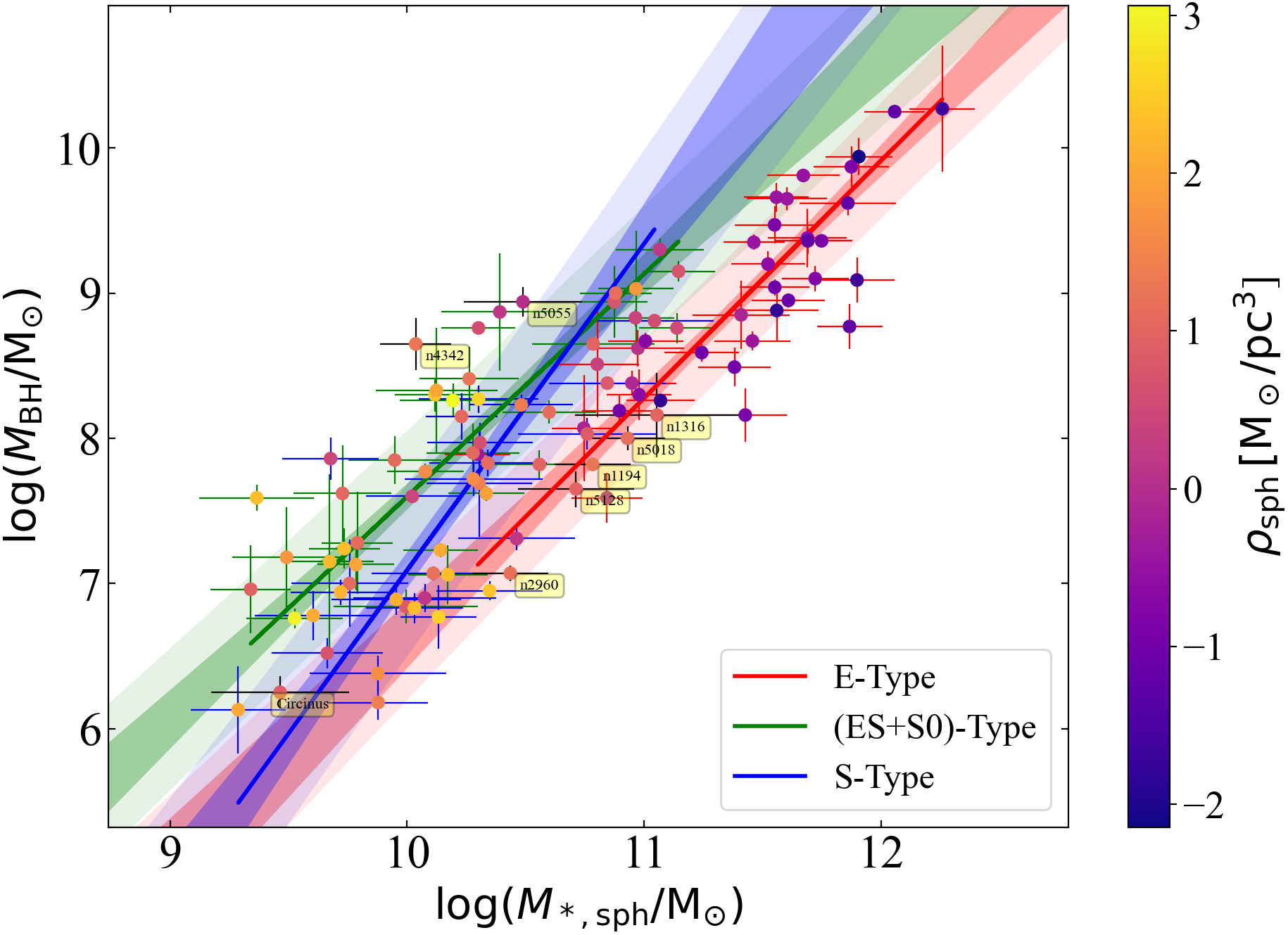}
\caption{Similar to the right hand-panel of 
  Fig.~\ref{Fig-M-R} but with a colour scale-bar on
  the right-hand side showing the spheroids' stellar density: 
  $M_{\rm *,sph}/[(4\pi/3)R^3_{\rm e,sph,eq}]$ [$M_\odot$/pc$^3$]. 
}
\label{Fig-M-M-rho}
\end{center}
\end{figure}

In Fig.~\ref{Fig-M-M-rho}, we show the stellar density of the ES and other
galaxy types in the $M_{\rm bh}$-$M_{\rm *,sph}$ diagram.  This figure is a
variant of Fig.~11 in \citet{2021arXiv211005037S}\footnote{Note:
  \citet[][see their Table~2]{2021arXiv211005037S} use the density within the
  internal half light radius and within a sphere of radius equal to the
  projected half-light radius.}, which showed the $M_{\rm bh}$-$\rho_{\rm
  e,sph}$ diagram for the E, ES/S0 and S galaxy
types. \citet{2021arXiv211005037S} additionally present results in terms of
the internal density and the projected density, aka `compactness', as a
function of the radius used to measure these enclosed densities.
Fig.~\ref{Fig-M-M-rho} shows that the most compact spheroids do not host the
most massive black holes.  Instead, such black holes are found in the most
diffuse spheroids, the giant E galaxies built from mergers.  The most diffuse
spheroids also tend to reside to right of the E galaxy sequence.


%

\subsection{Brightest cluster galaxies}\label{Sec_BCGs}

We already have key insight into the morphogenesis and ontogeny of spheroids.
For example, S0$+$S0 unions may create E galaxies, branching
off from the $M_{\rm bh}$-$M_{*,bulge}$ relation. 
%
%
Indeed, major dry mergers in which the large-scale angular momentum is erased
will produce elliptical galaxies \citep{2006ApJ...636L..81N}.
\citet{Graham:Sahu:22a} noted that a single S0 merger is sufficient to explain
the average offset of the elliptical galaxies in the $M_{\rm bh}$-$M_{\rm
  *,sph}$ diagram \citep{2019ApJ...876..155S}.  Some BCGs have likely formed
from the merger of E galaxies.  Such an equal mass merger would roughly double
both the black hole and spheroid mass, thereby forming an offshoot in the
scaling diagram following a trajectory with a slope of 1.

Some massive E galaxies may have suffered such extensive erosion by infalling
perturbers \citep{2010ApJ...725.1707G} and supermassive black holes
\citep{2005LRR.....8....8M} that it has fundamentally re-shaped the remnant
galaxy to yield a system well described by a low-$n$ S\'ersic profile.
Indeed, some cD galaxies at the centres of clusters are likely built by
multiple mergers, evidenced by multiple bright nuclei in the same system.
They are sometimes seen to consist of a low-$n$ spheroid surrounded by
intracluster light with a tendency to have an exponential light profile
\citep{2007MNRAS.378.1575S}.  Systems involving such a halo are referred to as
cD galaxies.
%

As \citet{Graham:Sahu:22a} illustrated, in terms of $M_{\rm bh}/M_{\rm *,sph}$
ratios, the E galaxies represent a population built by mergers.  The
super-linear $M_{\rm bh}$-$M_{\rm *,sph}$ relation observed for brightest
group galaxies (BGGs) and BCGs by \citet{2018ApJ...852..131B} is understood
here in terms of mergers which built the offset E galaxy sequence
\citep{2019ApJ...876..155S}.  By definition, they do not contain substantial
discs or significant rotation.  With BCGs and BGGs representing systems that
may have experienced the highest number of mergers, here we check if they
occupy a distinct region of the $M_{\rm bh}$-$M_{\rm *,sph}$ and $M_{\rm
  bh}$-$R_{\rm e,sph}$ diagram.

In Table~\ref{Table-BCG}, we identify the BCGs and BGGs in our sample.  If
they are a known cD galaxy, this is also reported.  Our imaging data may not
have always been deep/extended enough to capture the contribution from the
ICL. For four of the six cD galaxies, we did not detect the presence of the
ICL.  This non-detection is only a problem if there is an ICL component that
has influenced a portion of the outer-light profile included in the galaxy
decomposition work.  Within the $M_{\rm bh}$-$M_{\rm *,sph}$ diagram, we do
not detect a separation of BCG/BGG from the non-(BCG/BGG) elliptical galaxies.
While there is likely benefit to be had by acquiring and investigating new
deep images of the cD galaxies in our sample, such an undertaking is beyond
the intended scope of the present research project.

The galaxy with the largest directly measured black hole is Holm~15A, with an
adopted distance modulus of 37.02 mag.  Holm~15A is an intriguing galaxy which
was initially reported to have a vast 4.57$\pm$0.06~kpc depleted core
\citep{2014ApJ...795L..31L} which was later queried by
\citet{2015ApJ...807..136B} and \citet{2016ApJ...819...50M}. Although lacking
a noticeable downward break in its light profile, it is conceivable that this
galaxy has been so heavily eroded that what may have once been a high S\'ersic
index galaxy now resembles a low S\'ersic index system with no discernable
break in its light profile.  To distinguish this potential scenario from the
less severe core depletion seen in many big spheroids, we coin a new phrase
and refer to such dramatic reshaping as `galforming', in a somewhat similar
vein to the term terraforming.  Other brightest cluster galaxies, such as
NGC~5419 (Table~\ref{Table-BCG}), NGC~4874 in the Coma cluster and UGC~9799
(not in our sample), are also known to display this behaviour of an unusually
low S\'ersic index for their luminosity, embedded in a halo of intracluster light
\citep{2007MNRAS.378.1575S}.  NGC~4472 (M49), the brightest galaxy in
the Virgo~B subcluster, has also been reported to have an unexpectedly low
S\'ersic index \citep[][their
  Appendix~C]{2014MNRAS.444.2700D}.\footnote{According to
  \citet{1981rsac.book.....S}, NGC~4472 is an S0 galaxy. From photometry,
  \citet{2011MNRAS.418.1452L} fit a large-scale exponential function which
  dominates the light beyond $\sim$ 100$\arcsec$, although we wonder (but
  have not explored) if this might be a halo of ICL given that other works
  regard NGC~4472 as an E galaxy.}

\citet{2019ApJ...887..195M} report $M_{\rm bh} = (4.0\pm0.8) \times 10^{10}
M_\odot$ for Holm~15A.  Presumably, when measuring the brightness of BCGs, one
needs to exclude the ICL, which belongs to the cluster, rather than the
BCG. This was done by \citet{2015ApJ...807..136B}, who reported a CFHT $r$-band
magnitude of $\sim$13.8 mag (AB system).  Using $M_{\odot,r} = 4.64$ mag, and
adopting an $M_*/L_r$ ratio of 3$\pm$1, gives a stellar mass $M_{*,sph} =
4.2^{+1.4}_{-1.0} \times 10^{11}\,M_\odot$.  On the other hand,
\citet{2019ApJ...887..195M} include the ICL and adopt an $M_*/L_i$ ratio of
4.5$\pm$0.19 to report a galaxy+ICL stellar mass of $(25.0\pm6.4)\times
10^{11}\,M_\odot$, six times larger than the galaxy stellar mass reported by
\citet{2015ApJ...807..136B}.  While this places it smack on the super-linear
$M_{\rm bh}$-$M_{\rm *,sph}$ relation for elliptical galaxies, the inclusion
of the ICL appears questionable.

Finally, we note that KH13 excluded NGC 4889 (Coma) and NGC~3842 (Leo) due to
these galaxies' high black hole masses relative to the near-linear $M_{\rm
  bh}$-$M_{\rm *,sph}$ relation in KH13.  However, these galaxies do not have
an over-massive black hole relative to the morphology-dependent, super-linear
$M_{\rm bh}$-$M_{\rm *,sph}$ relations from \citet{Graham:Sahu:22a}, and we
do not exclude them.  KH13 also excluded NGC~1316, NGC~3607, NGC~4261 and
IC~4296 (Abell 3565-BCG) from their regression.  They also did not have
NGC~1275 and NGC~5419 in their sample.  As such, they only used 6 of the 14
galaxies listed in Table~\ref{Table-BCG}.

\begin{table*}
\centering
\caption{Brightest cluster galaxies, cD galaxies, and brightest group galaxies}
\label{Table-BCG}
\begin{tabular}{llcrlll}
\hline
Galaxy    &  Type   & ICL fit & $n_{\rm sph}$ & Ref   &  Morphology & Parent group/cluster. \\
\hline
IC 4296   &  BCG    &  no     &  3.8  & SGD19 &  Elliptical & Abell 3565 Cluster. \\ 
NGC~1275  &  BCG,cD &  no     &  4.3  & SGD19 &  Ellicular pec.\ (ES,e) & Perseus Cluster. \\
NGC 1316  &  BCG    &  no     &  1.8  & SG16  &  Lenticular pec. & aka Fornax A. Fornax Cluster. \\ 
NGC 1399  &  cD     &  yes    & 10.0  & Fig.~\ref{Fig-N821} & Elliptical pec. & 2nd brightest in Fornax Cluster. \\
NGC 3842  &  BCG    &  no     &  8.2  & SG16  &  Elliptical & Leo Cluster. \\ 
NGC 4472  &  cD     &  no     &  5.4  & SG16  &  Elliptical/S0? & aka M49 and Virgo B. 2nd brightest in Virgo Cluster. \\ 
NGC 4486  &  BCG,cD &  no     &  5.9  & SG16  &  Elliptical pec. &  aka M87 and Virgo A. Virgo Cluster. \\ 
NGC 4889  &  BCG,cD &  no     &  6.8  & SG16  &  Elliptical (rare E4) &  NGC~4874/4889 Coma Cluster. \\ 
NGC 5419  &  BCG,cD &  yes    &  2.6  & Fig.~\ref{Fig-N5419} & Elliptical & Abell S0753 Cluster. \\ 
NGC 7768  &  BCG    &  no     &  6.7  & SG16  &  Elliptical & Abell 2666 Cluster. \\ 
\hline 
IC 1459   &  BGG    &  no     &  7.0  & SGD19 &  Elliptical & IC 1459 Group (N=16). \\
NGC 0524  &  BGG    &  no     &  2.2  & SGD10 &  Lenticular & NGC 524 Group (N=16). \\ 
NGC 1332  &  BGG    &  no     &  3.7  & SGD19 &  Ellicular (ES,b) & NGC 1332 Group (N=22) in Eridanus Cluster. \\
NGC 1407  &  BGG    &  no     &  3.9  & SGD19 &  E with undigested cpt &  NGC~1407 Group (N=25) in Eridanus Cluster. \\
NGC 3031  &  BGG    &  no     &  3.5  & DGC19 &  Spiral (SAab) & aka M81. NGC~3031 Group (N=30). \\
NGC 3091  &  BGG    &  no     &  6.6  & SGD19 &  Elliptical & Hickson Compact Group No.\ 42. \\
NGC 3368  &  BGG    &  no     &  1.0  & DGC19 & Spiral (SABab) &  aka M96. M96 Group, aka Leo I Group. \\ 
NGC 3379  &  BGG    &  no     &  5.3 & SG16  &  Elliptical & aka M105. NGC 3379 Group (N=27). \\ 
NGC 3607  &  BGG    &  no     &  5.6  & SG16  &  Ellicular (ES,e) & NGC 3607 Group (N=31), Leo II Group, equal brightest \citep{2015AstBu..70....1K}. \\ 
NGC 3608  &  BGG    &  no     &  5.7  & SG16  &  Elliptical &  Leo II Group, equal brightest. \\ 
NGC 3627  &  BGG    &  no     &  2.1  & DGC19 &  Spiral (SABb) &  aka M66. N3627 Group (N=16) in Leo II Group. \\ 
NGC 3665  &  BGG    &  no     &  2.7  & SGD19 &  Lenticular & NGC 3665 Group (N=16). \\ 
NGC 3923  &  BGG    &  no     &  4.8  & SGD19 &  Elliptical with shells & NGC 3923 Group (N=23). \\
NGC 4151  &  BGG    &  no     &  1.9  & DGC19 & Spiral (SABab) & NGC 4151 Group (N=16). Binary BH? \citep{2012ApJ...759..118B}. \\ 
NGC 4258  &  BGG    &  no     &  2.6  & DGC19 & Spiral (SABbc) & aka M106. NGC 4258 Group (N=15). \\ 
NGC 4261  &  BGG    &  no     &  4.3  & SG16  &  Elliptical & NGC 4261 Group (N=87) \citep{1995ApJ...444..582D}. \\ 
NGC 4303  &  BGG    &  no     &  0.90 & DGC19 &  Spiral (SABbc) &  aka M61. NGC 4303 Group (N=23). \\ 
NGC 4501  &  BGG    &  no     &  2.8  & DGC19 &  Spiral (SAb) & aka M88. NGC 4501 Group (N=31). \\ 
NGC 4552  &  BGG    &  no     &  5.4  & SGD19 &  Elliptical & aka M89. NGC 4552 Group (N=12). \\ 
NGC 4594  &  BGG    &  no     &  4.2  & DGC19 &  Spiral (SAa) or ES,e  & aka M104. NGC 4594 Group (N=11). \\ 
NGC 4697  &  BGG    &  no     &  6.7  & SG16  &  Ellicular (ES,e) & NGC 4697 Group (N=37). \\ 
NGC 5018  &  BGG    &  no     &  2.5  & SGD19 &  Merger S0 pec. & NGC 5018 Group (N=9). \citep{1996AnA...314..357H}. \\
NGC 5128  &  BGG    &  no     &  2.2  & SG16  &  Merger S0 pec. & aka Cen-A. NGC 5128 Group (N=15). \\
NGC 5846  &  BGG    &  no     &  5.7  & SGD19 &  Elliptical &  NGC 5846 Group (N=74). \\
NGC 7582  &  BGG    &  no     &  2.2  & DGC19 &  Spiral (SBab) & NGC 7582 Group (N=13). \\
\hline 
\end{tabular}

Column~2 indicates if the galaxy is the brightest in a group or cluster or a cD galaxy. 
Column~3 indicates if we (thought we needed to) fit for intracluster light, which is
dependent on the surface brightness limit to which a galaxy is modelled. 
Column~4 lists the equivalent-axis S\'ersic index. 
Column~5 provides the reference for the galaxy decomposition: 
SGD19 = \citet{2019ApJ...876..155S}; SG16 = \citet{2016ApJS..222...10S}. 
Column~6 provides the galaxy morphology, and 
Column~7 gives the group/cluster identification
\citep{2004MNRAS.350.1511O, 2011MNRAS.412.2498M} if the membership count $N$
is $\gtrsim$10, along with an alternate galaxy name if it is also known as
(aka) something else. 
\end{table*}

\subsubsection{Exclusions}\label{Sec_exc}

As noted above, KH13 excluded many BCGs from their regressions while we find
that BCGs can naturally be included and explained in terms of mergers and
morphology-dependent black hole scaling relations.  In fact, 
KH13 excluded more than one-third (17/45) of the 'Elliptical galaxies'' in their set.
Six of these 17 (plus one S0 galaxy) appear to have been excluded because KH13
reported tentative evidence
for the BH masses to be low in systems for which the black hole mass
was measured using ionised emission lines but for which the width of the emission line
was not taken into account.  
For two (NGC~4261 and NGC~7052), we have new BH mass estimates from
the literature, as reported in \citet{Graham:Sahu:22a}. 
For NGC~4459 \citep{2011MNRAS.414.2923K}, KH13 considered the galaxy to be an
E rather than an S0 galaxy and consequently over-estimated the spheroid
brightness.  Adjusting for this, NGC~4459 is not an outlier in the $M_{\rm
  bh}$-$M_{\rm *,sph}$ diagram, nor is the S0 galaxy NGC~4596, which KH13
recognised as such.  The inclusion or inclusion of these two S0 galaxies has
no significant impact on our results.  This leaves Abell 1836-BCG (not in our
sample), Abell 1836-BCG (=IC~4296, kept in our sample) and the giant
elliptical galaxy NGC~6251, which we also keep with its active Seyfert 2
nucleus and a dusty nuclear disc \citep{1997ApJ...486L..91C}.  NGC~6251
resides in the $M_{\rm bh}$-$M_{\rm *,sph}$ diagram near NGC~1399 (the second
brightest galaxy of the Fornax cluster).  Given the cluster-centric location
and high galaxy stellar mass of NGC~1399, it is highly plausible that it was
built from a significant merger involving elliptical galaxies.  As such, it
can be expected to reside to the right of the distribution of elliptical
galaxies built from the merger of lenticular galaxies.  There is no additional
folding in of disc mass in E$+$E mergers; as such, they will migrate along a
line with a slope of unity in the $M_{\rm bh}$-$M_{\rm *,sph}$ diagram,
thereby moving off and to the right of the steeper super-linear relation
defined by the ensemble of elliptical galaxies.  This assumes that most E
galaxies in the ensemble are not the central dominant galaxy in a cluster and
have thus presumably not experienced quite as active merger histories.
Additional notes on galaxies that we have included but which KH13 excluded can
be found in Appendix~\ref{App_notes}.

For a century, astronomers have inadvertently overlooked, and thus excluded in
their analysis, the presence of discs in ETGs.  KH13 is no different and
regards eight galaxies containing discs as E galaxies.  In addition to
NGC~4459 noted above, plus the five late-stage mergers mentioned in
Section~\ref{Sec_wet_merger}, KH13 also treated the S0 galaxies NGC~1374
\citep{2019ApJ...876..155S} and NGC~2778 \citep{2011MNRAS.414.2923K,
  2016ApJS..222...10S} as E galaxies, overestimating the spheroid mass by a
factor of $\sim$2 and $\sim$4, respectively.  KH13 additionally treat the ES
galaxies NGC~1332, NGC~3377, NGC~3607, NGC~4473, NGC~4697, NGC~5845 and
NGC~6861 (Table~\ref{TableESeESb}) as E galaxies, and thus they assign
$\sim$5--25 per cent too much flux/mass to the spheroidal component of these
seven galaxies.
%

\begin{figure*}
\begin{center}
\includegraphics[trim=0.0cm 0cm 0.0cm 0cm, width=0.8\textwidth, angle=0]{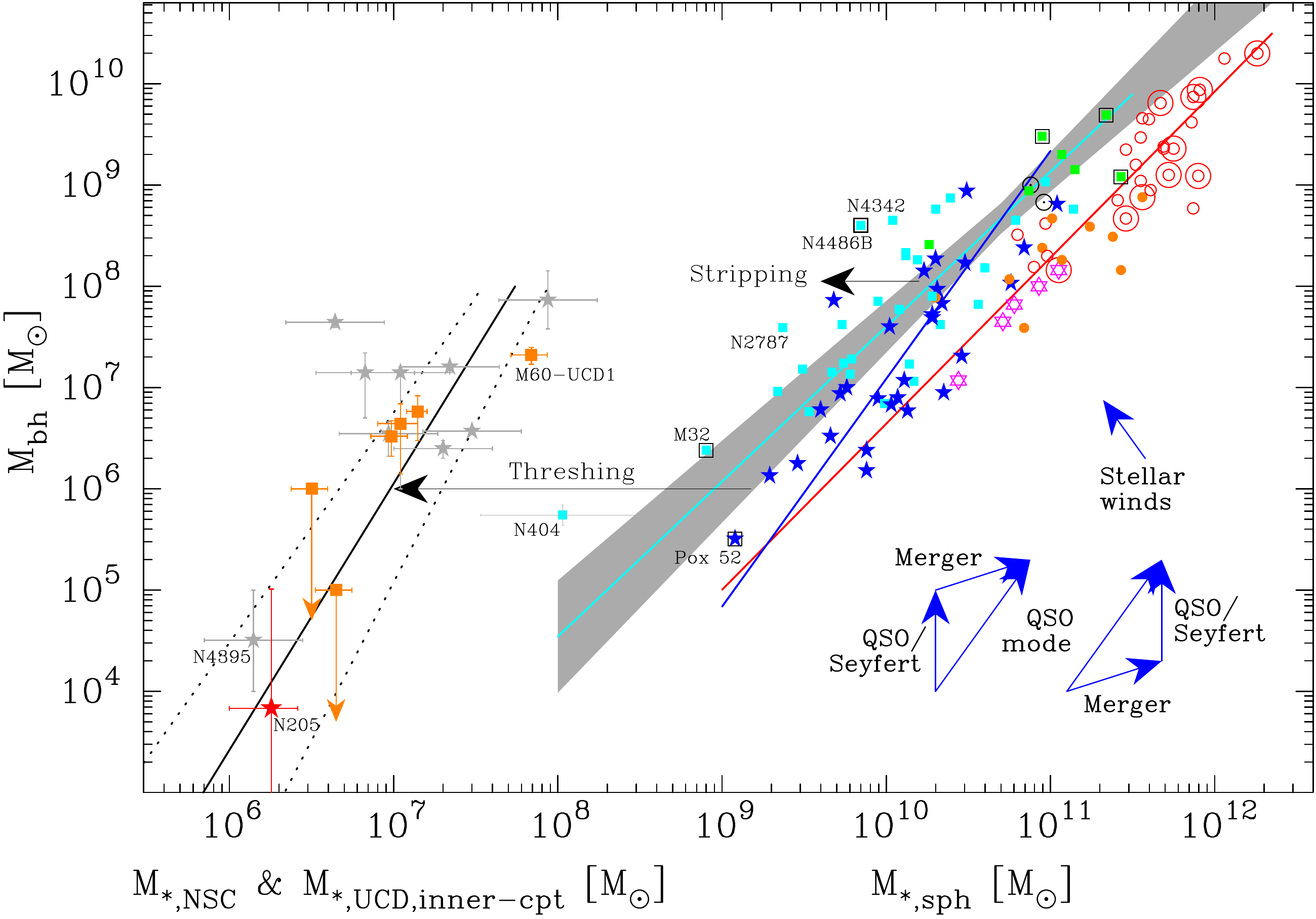}
\caption{Expansion of the right-hand panel of Fig.~\ref{Fig-M-R-2} to include
  ultracompact dwarf galaxies (orange squares) and nuclear star clusters
  (NSCs: grey
  stars) with $M_{\rm bh}<10^8$ M$_\odot$, along with the spheroids. The vertical
  axis displays the black hole mass while the horizontal axis displays the
  stellar mass of the above three systems. 
The remaining symbols have the same meaning as in Figure~\ref{Fig-M-R-2}. 
The non-Spitzer 
  galaxies NGC~4486B \citep{2016ApJ...818...47S}, plus M32 and Pox~52
  (Fig.~\ref{Fig-M-R}), have been overlaid on the figure.  The tentative $M_{\rm
  bh}$-$M_{\rm nsc}$ relation proffered by \citet{2016IAUS..312..269G} and
updated in \citet{2020MNRAS.492.3263G}, which holds for 
$M_{\rm bh}\lesssim 10^8$ M$_\odot$, is also shown.  
}
\label{Fig-M-M-strip}
\end{center}
\end{figure*}

\subsection{Stripped and threshed galaxies}\label{Sec_Strip}

Soon after William Parsons\footnote{William Parsons was the 3$^{\rm rd}$ Earl
  of Rosse, Ireland, and is often referred to as Lord Rosse.} 
discovered the first spiral nebulae \citep{1850b_Rosse, 1850RSPT..140..499R},
the tidal theory of \citet{Roche:1850} explained how a massive 
galaxy could gravitationally strip stars from a neighbouring galaxy.
\citet{1852AJ......2...95A} reasoned that the relatively stronger (weaker)
gravitational force on the facing (opposing) sides of the galaxy 
would lead to two protrusions 
which would be drawn out into spiral-like arms if the galaxy rotated. 
Indeed, tidal arms and tails have since been modelled
\citep[e.g.,][]{1972ApJ...178..623T}, and are well known
\citep[e.g.,][]{2000ASPC..218..349B, 2006AJ....131..261J}.  It is the
ongoing tidal-stripping which is thought to eventually pare away a galaxy's
disc, rather than the outskirts of a large E galaxy,  
 to create a more dominant bulge \citep{2001ApJ...557L..39B}, as seen in the `compact
elliptical' (cE) galaxies.  
If the bulge is not stripped, ignoring any black hole growth or spheroid
mass loss due to stellar winds, the system will remain on the S0 $M_{\rm
  bh}$-$M_{\rm *,sph}$ relation shown in Fig.~\ref{Fig-M-R}. 
%
%
Ongoing stripping may pare away the bulge, leaving
the more resilient nuclear star cluster to become a UCD 
galaxy, in a process referred to as threshing 
\citep[e.g.,][and references therein]{2001ApJ...552L.105B, 2020MNRAS.492.3263G}. 
The migration from these processes is illustrated in Fig.~\ref{Fig-M-M-strip}, 
including the compact elliptical galaxies M32 and NGC~4486B, the
stripped galaxy NGC~4342, and several UCDs. This pattern has been seen before,
for example, \citet[][their Fig.~7]{2021MNRAS.506.4702F}.


\citet{2010MNRAS.401.1770V} report $M_{\rm bh} =
(2.4\pm1.0)\times10^6\,M_\odot$ for M32.  \citet[][their
  Table~1]{2016ApJ...818...47S} report $\log\,M_{\rm *,sph} = 8.627\pm0.022$
based on $\log(L_{V,M32}/L_{V,\odot})=8.572$ from \citet{1998AJ....115.2285M},
along with a $V$-band Galactic extinction of 0.206 mag, and the use of
$M_*/L_V=1 M_\odot/L_\odot$ rather than $M_*/L_V = 2.18 M_\odot/L_\odot$ as
reported by \citet{1998AJ....115.2285M}.  We depart from this mass estimate in
two ways.  First, the adopted mass-to-light ratio is probably too small for
the ancient stellar population in M32 \citep{2009MNRAS.396..624C}, which has
$B-V \approx0.95$ \citep{1991rc3..book.....D}.  The models of
\citet{2022AJ....163..154S} suggest $M_*/L_V = 5.0$ $M_\odot/L_\odot$, which
we reduce by $10^{-0.1}$, i.e., 0.1 dex, to give $M_*/L_V = 4.0$
$M_\odot/L_\odot$ for a \citet{2002Sci...295...82K} initial mass function.
Second, M32 is known to have a three-component structure
\citep{2009MNRAS.397.2148G}, including a weak disc \citep{2002ApJ...568L..13G}
that was likely eroded by tidal-stripping \citep{1962AJ.....67..471K,
  1973ApJ...181...27K, 1965AJ.....70T.689R, 1973ApJ...179..423F,
  2001ApJ...557L..39B}.  As such, we reduce the galaxy luminosity by a factor
of 0.62 \citep{2002ApJ...568L..13G} to obtain the spheroid luminosity.  In so
doing, we obtain a Galactic extinction corrected \citep[$A_V=0.17$:
][]{2011ApJ...737..103S} stellar mass for the spheroidal component of M32
equal to $1.08\times10^9\,M_\odot$.
This is just 35 per cent higher than the preferred derivation provided in
\citet{Graham:Sahu:22a}, which we use here. 

Another well known cE galaxy is  NGC~4486B, for which
\citet{1998AJ....115.2285M} report 
$\log(L_{V,gal}/L_{V,\odot})=8.960$ and $M_*/L_V=3.6\pm0.4$, giving
$\log(M_{\rm *,gal}/M_\odot) = 9.516$.  
\citet{2016ApJ...818...47S} report a higher value of 
$\log(M_{\rm *,gal}/M_\odot) = 9.847\pm0.027$, and 
$\log(M_{\rm bh}/M_\odot) = 8.602\pm0.024$, which we plot in
Fig.~\ref{Fig-M-M-strip} along with M32. 
%
%
While M32 may be the result of stripping much of the disc from an S0 galaxy
while leaving the spheroidal component largely intact, NGC~4486B may have
additionally had some of the spheroidal component removed.\footnote{As
  revealed in \citet[][their Fig.~11]{2019MNRAS.484..794G}, IC~3653 (not in
  our sample) is another galaxy which might be expected to have an elevated
  $M_{\rm bh}/M_{\rm *,sph}$ ratio due to tidal stripping if its bulge
  component has been depleted.}
%
%
However, confirming this would require further investigation.  For example,
while the S0 galaxy NGC~2787 appears  to the left 
of the (cold gas)-poor S0 sequence in the $M_{\rm bh}$-$M_{\rm *,sph}$ diagram,
similar to NGC~4486B, it is not regarded as a tidally-stripped galaxy. 


Although not a cE galaxy, the S0 galaxy NGC~404 also resides at the low-mass
end of our scaling relations.  NGC~404 is marginally consistent with the black
hole scaling relations for the predominantly non-(star-forming) S0 galaxies,
as seen in the lower left of the panels in Fig.~\ref{Fig-M-R}.  While most
of the stars in NGC~404 are old, it does contain cold gas in its disc but only
displays a declining tail of star formation in its outer disc
\citep{2010ApJ...716...71W, 2013ApJ...772L..23B}.  If NGC~404, or the dwarf
ETG Pox~52 \citep{2008ApJ...686..892T}\footnote{Pox~52 appears to have an AGN
  plus a two-component stellar structure.  It may be a dwarf S0 galaxy with a
  disc and bulge, with the unsubtracted portion of the bulge visible in the
  residual image of \citet[][their Fig.~11.50]{2017ApJS..232...21K}.}, were
to merge with a larger bulgeless LTG, the remnant might resemble the
post-merger spiral galaxy NGC~4424 \citep{2021ApJ...923..146G}, which can be
appreciated from Fig.~\ref{Fig-M-R-3}.  Although, we do caution that more,
and reliable, data is needed at low masses before a secure picture can be
established in the low-mass regime.
\citet{2017ApJ...839L..13S} has argued that the presence of IMBHs 
may be a common feature of old dwarf galaxies. Unlike in
low-mass LTGs, IMBHs may be hard to spot due to their expectedly lower
Eddington ratios, possibly explaining the lower detection rates in ETGs than
LTGs \citep{2019MNRAS.484..794G, 2019MNRAS.484..814G}.

\begin{figure}
\begin{center}
\includegraphics[trim=0.0cm 0cm 0.0cm 0cm, width=1.0\columnwidth, angle=0]{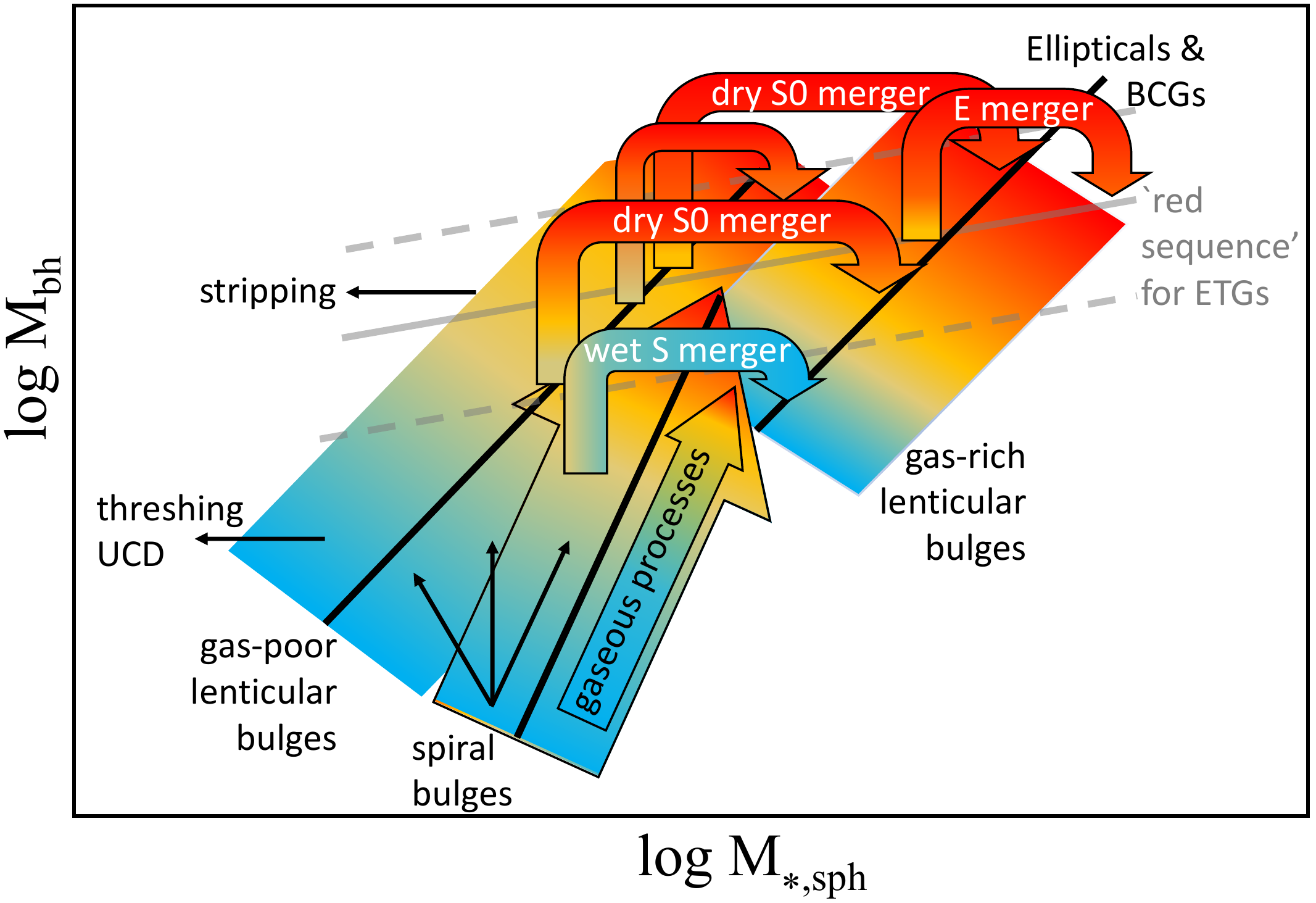}
\caption{Adapted from Fig.~A8 in \citet{Graham:Sahu:22a}.  Here we unveil
  further drivers of galaxy growth in the $M_{\rm bh}$-$M_{\rm *,sph}$
  diagram.  The three upward-pointing black arrows show potential pathways for
  gas-rich disc galaxies.  From left to right, they are: stellar winds which
  cause stellar mass loss in the spheroid but can slowly feed the central
  black hole \citep{1991ApJ...376..380C, 2006ApJ...640..143S}; rapid AGN
  accretion of cold gas fuelling Seyfert/Quasar growth (and possibly disc
  growth but not spheroid growth); and collective AGN/spheroid growth from
  `cold mode', aka `quasar mode', activity \citep{2000MNRAS.311..576K}. This
  schematic implicitly assumes a past presence, when the mergers occurred, of
  S and S0 disc galaxies distributed roughly along the indicated trend lines
  defined by today's population.  }
\label{Fig-schematic_3}
\end{center}
\end{figure}

\section{Summary}\label{Sec_Sum}


In Figures~\ref{Fig-M-M-strip} and \ref{Fig-schematic_3}, we have
tried to summarise key movements in the $M_{\rm bh}$-$M_{\rm sph}$ diagram.
As noted in \citet{Graham:Sahu:22a}, elliptical galaxies built from the merger
of (two) lenticular galaxies can produce a notable jump in this diagram due to
folding all of the lenticular galaxies' disc stars into the new elliptical
galaxy.  As was also revealed in \citet{Graham:Sahu:22a}, these jumps are
mirrored in the $M_{\rm bh}$-$R_{\rm e,sph}$ diagram (see
Fig.~\ref{Fig-M-R}).  Obviously, but perhaps worth stating, is that if these
mergers occurred long ago, then lenticular disc galaxies, and their offset
sequence from the elliptical galaxies in the scaling diagram, existed long
ago.  This has implications for observation-based evolutionary studies
\citep[e.g.,][]{2011ApJ...742..107B, 2020ApJ...888...37D}, which would benefit from knowledge of the
high-$z$ galaxies' morphological type beyond simply ETG or LTG.

We looked for the signature of E$+$E mergers by checking for a displaced
population of brightest cluster (and group) galaxies relative to the general
distribution of elliptical galaxies, but we did not observe a clear signal
(Fig.~\ref{Fig-M-R-2}).  That is, our sample of BGGs/BCGs largely overlap
with the location of non-(BGGs/BCGs) elliptical galaxies in the $M_{\rm 
  bh}$-$M_{\rm sph}$ diagram. This may reflect that not all BGGs/BCGs are
built from a major E$+$E collision.  

We find tentative evidence for two types of ellicular galaxy: ancient compact
systems which have not accreted a large-scale disc and less dense systems
likely built by a substantial merger event (see Fig.~\ref{Fig-M-M-rho}.  We
also observe lenticular galaxies with partially depleted cores, likely built
by a dry S0 galaxy merger, near the top of the S0 sequence, and relatively
(cold gas)-rich lenticular galaxies built from a wet merger event, near the
bottom of the merger-built elliptical galaxy sequence.  Neither the (cold
gas)-poor lenticular sequence nor the elliptical galaxy sequence are
linear. As noted in \citet{Graham:Sahu:22a}, they have a superlinear, or
near-quadratic, slope of $\sim$1.6.  These ETGs with discs are examples of
mergers in which some of the orbital angular momentum from the pre-merged
binary remains.  Indeed, mergers which do not cancel the orbital angular
momentum can result in the formation of S0 and ES galaxies
\citep{1992ApJ...393..484B, 2006MNRAS.372..839N}.  Figures~\ref{Fig-M-R-3} and
\ref{Fig-M-R-4} 
present the $M_{\rm bh}$-$M_{\rm sph}$ diagram as a function of galaxy
morphology, revealing the evolutionary pathways of galaxies and their black
holes, mirroring the morphology-dependent sequences in the $M_{\rm
  bh}$-$M_{\rm sph}$ diagram.

%
%

Building upon the Darwinian concepts of punctuated equilibrium and gradualism
--- introduced in \citet{Graham:Sahu:22a} to help explain the two dominant sequences
seen in the $M_{\rm bh}$-$M_{\rm *,sph}$ diagram --- 
we start to present the beginning of what might become something of a 
phylogenetic tree to help bring out the additional information in this diagram. 
Although still some way from offering a complete picture of spheroid/(black
hole) evolution, akin to 
how the Hertzsprung-Russell diagram explains stars, the evolutionary paths shown here offer
a significantly increased understanding beyond the notion of a (near-linear)
$M_{\rm bh}$-$M_{\rm *,sph}$ relation with disjoint populations of low- and
high-mass spheroids having under- and over-massive black holes, respectively. 
The non-linear, morphology-dependent 
relations represent a markedly different picture to that based on
the idea of `classical bulges' versus `pseudobulges' \citep[e.g.,][their
  Fig.~13]{2008MNRAS.386.2242H, 2013ARA&A..51..511K, 2016ApJ...818...47S}
dictating the demographics within the black hole scaling diagrams. They also
represent a substantial advancement over the `red sequence' for ETGs
presented in \citep{2016ApJ...817...21S} along with the `blue sequence' for LTGs.
Our holistic approach also reminds us that some lenticular galaxies are not
faded spiral galaxies but have origins tied to (wet and dry) mergers. 
However, suggestions surrounding the environmental impact on discs, 
presented at the end of Section~\ref{Sec_wet_merger}, 
for the offset behaviour of ES,b 
and some S0 galaxies in the $M_{\rm bh}$-$M_{\rm *,gal}$ diagram, would
benefit from further investigation.

By combining our results with detailed morphological information for galaxies
with reverberation-mapped AGN \citep[e.g.,][]{2009ApJ...694L.166B,
  2017FrASS...4...12I, 2021ApJ...921...36B}, we can revisit and improve
derivations of the virial $f$-factor \citep{1993PASP..105..247P} used to
convert AGN virial masses into black hole masses.  The morphology-dependence
seen in Fig.~\ref{Fig-M-R-3} will enable improved estimates of black hole
masses in galaxies for which the gravitational influence of the black hole
cannot be spatially resolved, including systems displaying tidal disruption
events \citep[TDEs:][]{1988Natur.333..523R, 2009MNRAS.400.2070S, 2015JHEAp...7..148K}.
Furthermore, the new, all-encompassing, merger-built sequence in the $M_{\rm bh}$-$M_{\rm
  *,gal}$ diagram, to which S galaxies also roughly adhere to, may
prove helpful for pursuing long-wavelength gravitational waves generated
through the coalescence of massive black holes
\citep[e.g.,][]{2012A&A...542A.102M, 2013Sci...342..334S, 2019MNRAS.488..401C}.

\section*{Acknowledgements}

This research was supported under the Australian Research Council's funding
scheme DP17012923.
Part of this research was conducted within the Australian Research Council's
  Centre of Excellence for Gravitational Wave Discovery (OzGrav) through
  project number CE170100004.
This research has used the NASA/IPAC Extragalactic Database (NED) and
the NASA/IPAC Infrared Science Archive. 
We used the {\sc Rstan} package available at \url{https://mc-stan.org/}. 
We also used python packages {\sc NumPy} \citep{harris2020array}, 
{\sc Matplotlib} \citep{Hunter:2007}, and {\sc SciPy} \citep{2020SciPy-NMeth}. 

We acknowledge having borrowed terms from our colleagues in the biological
sciences, such as `natural selection' from Darwin (1859) 
and `phylogeny' and `ontogeny' from Haeckel (1866).

\section{Data Availability}

The imaging data underlying this article are available in the NASA/IPAC Infrared
Science Archive.  The masses and sizes first displayed in Fig.~\ref{Fig-M-R} are
tabulated in \citet{Graham:Sahu:22a}.

\bibliographystyle{mnras}
\bibliography{Paper-BH-mass2}{}

\appendix

\section{Appendix}
\label{Apdx1}

\subsection{New multicomponent decompositions of the galaxy light}


We provide new decompositions of the galaxy light for seven galaxies imaged
with the Spitzer Space Telescope.  The data and process were described in our past
papers \citep{2019ApJ...873...85D, 2019ApJ...876..155S}.  The new IRAF task
{\sc Isofit} \citep{2015ApJ...810..120C} was used to extract the 
equivalent-axis light profile, which we modelled with {\sc Profiler}
\citep{2016PASA...33...62C}.  The results are shown here, in AB mag,
while the spheroid and galaxy masses, and the spheroid size, are 
reported in \citet[][their Table~1]{Graham:Sahu:22a}.

\subsubsection{NGC~821}

\citet{2016ApJ...817...21S} modelled the major- and equivalent-axis light
profiles independently of each other.  On the one hand, this offered insight
into some of the systematic errors involved in the quantitative structural
analysis of galaxies.  On the other hand, the reader was occasionally left
uncertain which decomposition was optimal. In general, fits to the major-axis
are desirable for obtaining the disc scalelength, while fits to the
equivalent-axis are preferable for obtaining the spheroid size, $R_{\rm e,sph}$.  
This was the case for NGC~821.  
\citet{2016ApJ...817...21S} additionally modelled this galaxy, and some 60
others with directly measured black hole masses, using the two-dimensional
fitting code {\sc Imfit} \citep{2015ApJ...799..226E}. However, as explained
there \citep[see also][for the pros and cons of two-dimensional modelling]{2015ApJ...810..120C}, it was found 
that the two-dimensional modelling was not as reliable as our preferred approach
which captures the (symmetrical) two-dimensional information in a series of one-dimensional 
profiles, including the ellipticity profile, the position angle profile, and Fourier harmonic
terms describing deviations in the isophotes from pure ellipses. This approach
allows for, for example, position angle twists in triaxial bulges.  

\begin{figure}
\begin{center}
\includegraphics[trim=0.0cm 0cm 0.0cm 0cm, width=\columnwidth,angle=0]{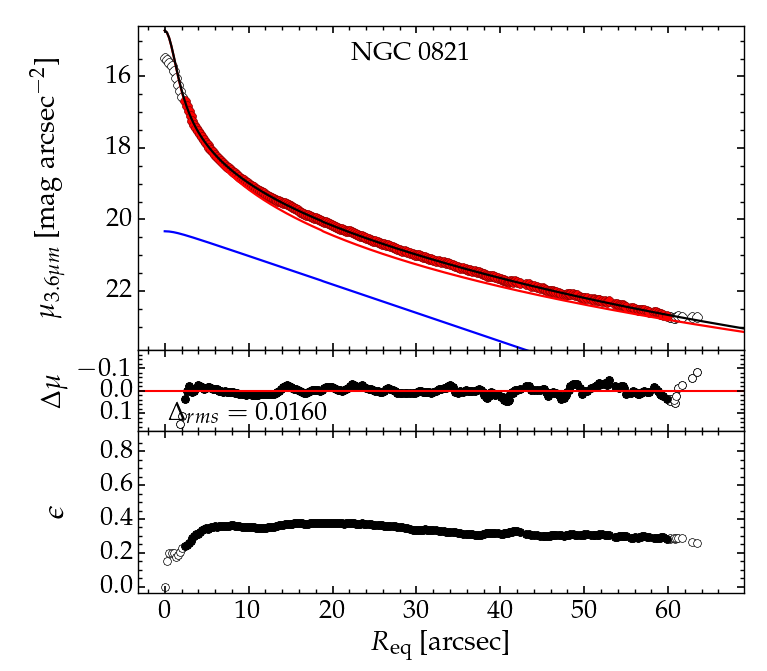}
\caption{The `equivalent-axis' light profile for NGC~821 (AB mag) fit with a S\'ersic
  spheroid (red) plus an intermediate-scale exponential disc (blue). Inclusion of the disc
  gives a spheroid magnitude 0.1 mag fainter than the galaxy magnitude.}
\label{Fig-N821}
\end{center}
\end{figure}

We use the {\sc iraf} task
{\sc Isofit} \citep{2015ApJ...810..120C} to analyse a 3.6~$\mu$m image of
NGC~821 taken from the Spitzer Heritage Archive \citep[SHA:
][]{2010SPIE.7737E..16W}\footnote{\url{https://sha.ipac.caltech.edu/}}.  We
obtain and show in Fig.~\ref{Fig-N821} an equivalent-axis light profile and
decomposition similar to that presented in \citet{2016ApJ...817...21S}.  Our
spheroid parameters from the equivalent-axis light profile are: $R_{\rm
  e}=31\arcsec.1\pm0.8$, $\mu_{\rm e}=21.33\pm0.05$ (AB) and $n=5.19\pm0.05$. 
Included are the formal/random errors determined from the decomposition
software \citep[Profiler][]{2016PASA...33...62C}.  Systematic errors can be
larger, and we, therefore, adopt the (quality of fit) grading scheme
established by \citet{2016ApJ...817...21S} and tweaked in
\citet{2019ApJ...876..155S}.  In doing so, we report an apparent magnitude
$m_{\rm sph}=10.34\pm0.20$ mag (AB) for the dominant spheroidal component of
NGC~821.\footnote{For reference,
\citet{2016ApJ...817...21S} obtained (7.85+2.76=) 10.61 mag (AB) from the
one-dimensional approach and (7.78+2.76=) 10.54 mag (AB) from the
two-dimensional approach. This is 0.20--0.27 mag fainter than found here 
because our more extended light profile yielded a fainter disc.}  
The galaxy magnitude is 10.24 mag (AB).

\subsubsection{NGC~1320}

\begin{figure*}
\begin{center}
\includegraphics[trim=0.0cm 0cm 0.0cm 0cm, height=0.3\textwidth,
  angle=0]{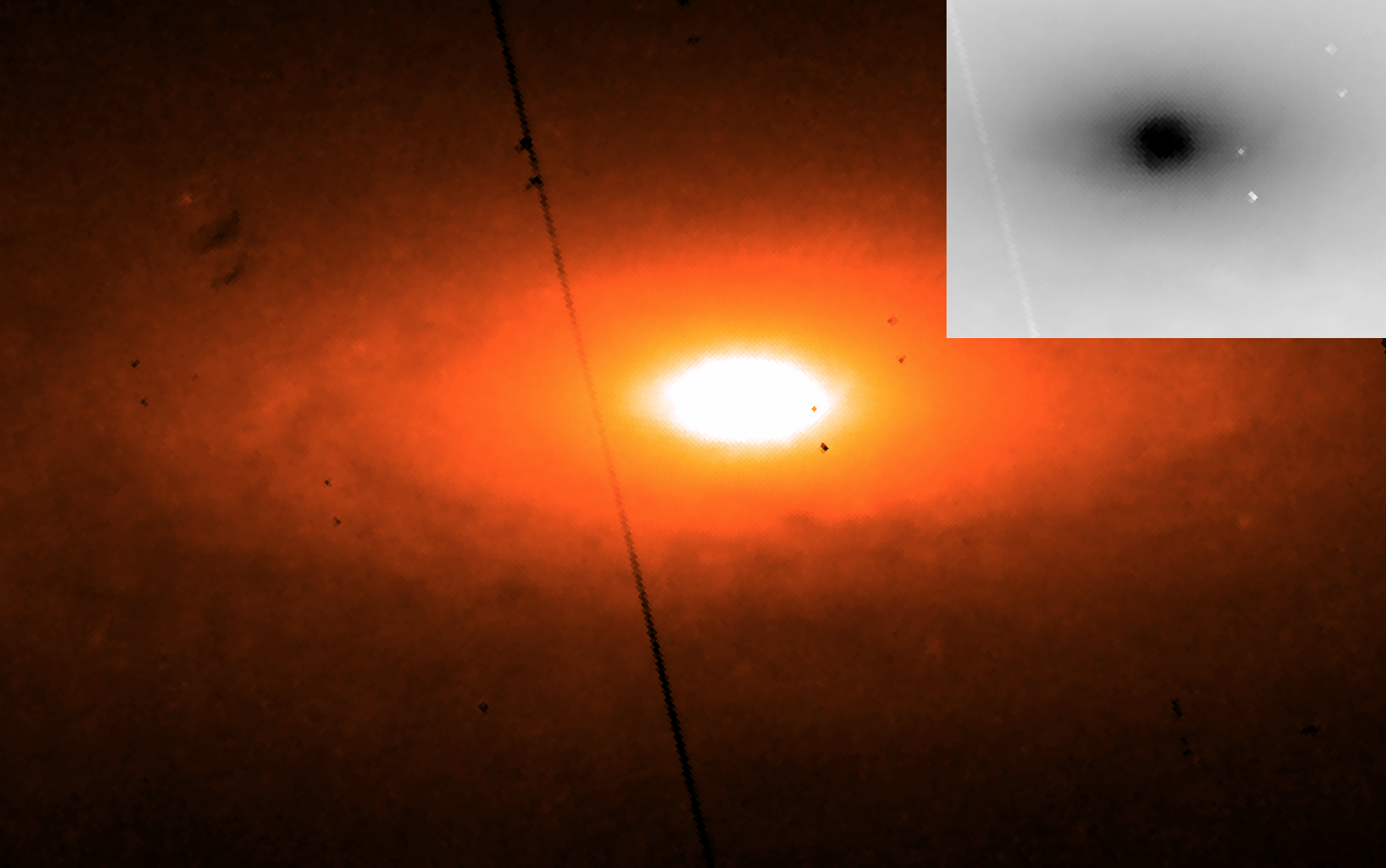}
\includegraphics[trim=0.0cm 0cm 0.0cm 0cm, height=0.32\textwidth,
  angle=0]{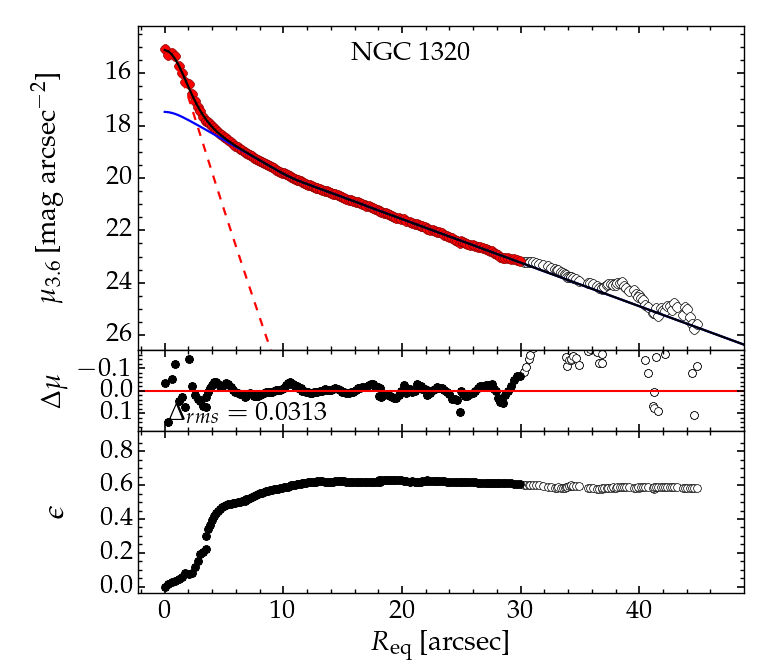}
\caption{ Left: Drizzled F160W/NIC2/NICMOS/HST image of NGC~1320, courtesy of
  the Hubble Legacy Archive (HLA) and HST Observing Proposal ID 7330 (PI:
  J.\ S.\ Mulchaey).  The pixel scale is $\sim$0.659 times the native 0.0757
  arcseconds pixel scale.  The primary field-of-view shown here is roughly
  $10.7\times17.1$ arcseconds. Zooming in reveals a spheroidal bulge
  surrounded by an elongated disc-like structure unlikely to be a bar.  Right:
  Decomposition of the Spitzer 3.6~$\mu$m light profile (equivalent-axis, aka
  geometric-mean axis) obtained from a larger field-of-view, involving a
  S\'ersic-bulge (red) plus an (anti-truncated)-disc (blue) with a bend at
  $\sim$8$\arcsec$, or instead, a disc with a brightened inner region.}
\label{Fig-N1320}
\end{center}
\end{figure*}

In Fig.~\ref{Fig-N1320}, we have remodelled NGC~1320, paying attention to
the rapid rise in the ellipticity profile from $\sim$0 to $\sim$0.5 over
2$\arcsec$ to 6$\arcsec$ \citep{2019ApJ...873...85D}.  This does not appear to
be the result of a bar in the image, nor does the light profile display the
characteristic bump arising from the light at the end of a bar.  Instead, this
elongated feature seems more associated with the disc, which has an
observed outer ellipticity of $\sim$0.6 due to its inclination angle relative
to our sight line.  However, in passing, we speculate that we may have
discovered an example of an early formation phase of a bar ($\lesssim
6\arcsec$), although, if this were the case, then perhaps the observed
ellipticity of this feature should be $\sim$0.7 rather than $\sim$0.5.  We
measure an apparent bulge magnitude, at 3.6~$\mu$m, of 12.99$\pm$0.25 (AB
mag).  The galaxy's apparent magnitude is 11.58 (AB mag).
%

\subsubsection{NGC~1399}

\begin{figure}
\begin{center}
\includegraphics[trim=0.0cm 0cm 0.0cm 0cm, width=\columnwidth, angle=0]{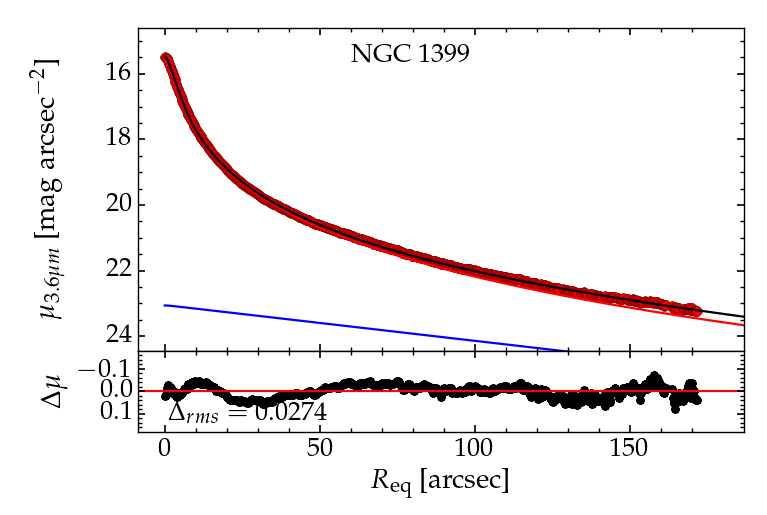}
\caption{Structural decomposition of NGC~1399 into galaxy (red) plus ICL
  (blue).  The equivalent-axis S\'ersic index of the galaxy is 7.0 and $R_{\rm
    e,sph,eq}=62\arcsec$.  The exponential model has a scalelength of
  100$\arcsec$, and it is apparent that a deeper exposure is required to
  sample the ICL surrounding this galaxy properly.}
\label{Fig-N1399}
\end{center}
\end{figure}

NGC~1399 is the cD galaxy in the Fornax cluster.  \citet{2013AJ....146..160R}
fit the starlight with a core-S\'ersic function for the central galaxy and a
S\'ersic plus exponential function for the intracluster light.  They
found an index $n=1.33$ for the S\'ersic function and a scalelength of over
1000$\arcsec$ for their exponential model, effectively mimicking a
sky-subtraction adjustment.  This result broadly resonates with
\citet{2007MNRAS.378.1575S}, who observed a preference for the ICL to be well
described with a (single) exponential function, i.e., an $n=1$ S\'ersic model.
%
%
Here, we adopt a simple approach by fitting a core-S\'ersic model for the
galaxy plus an exponential model for the ICL.  The apparent magnitude of the
galaxy\footnote{Adding our ICL component brightens the total magnitude by just
  0.10 mag.} is found to be 8.58 mag (AB) in the 3.6~$\mu$m band.  The
geometric mean-axis light profile is shown in Fig.~\ref{Fig-N1399}.

In passing, we note that there may be more to this galaxy.  We have not
sampled any ICL-dominated radii, and \citet[their
  Fig.~37]{2016ApJS..222...10S} thought there might be a faint inner
(undigested or unrelaxed) component which they modelled with an exponential
function.

\subsubsection{NGC~2787}

\begin{figure}
\begin{center}
\includegraphics[trim=0.0cm 0cm 0.0cm 0cm, width=\columnwidth,
  angle=0]{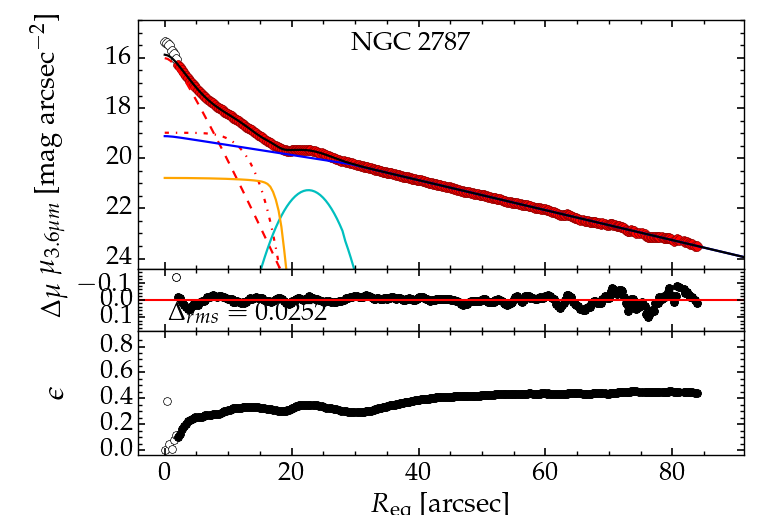}
\caption{Structural decomposition of NGC~2787.  The inner couple of arcseconds
  has been excluded to avoid (most of) the nuclear stellar rings \citep{2010MNRAS.402.2462C}
or nuclear stellar disc
  \citep{2010MNRAS.407..969L}.  The decomposition includes a S\'ersic bulge
  (red dashed), low-$n$ S\'ersic barlens (red dash-dot), Ferrers bar
  (orange), Gaussian-ring ansae (cyan) and truncated (bend at
  $R_{\rm eq}=29\arcsec.4$) exponential disc (blue).}
\label{Fig-N2787}
\end{center}
\end{figure}


NGC~2787 is the S0 galaxy with the lowest spheroid mass in our sample.  We
considered it prudent to revisit the decomposition.  However, with a bar, 
ansae at the ends of the bar, and a prominent barlens/(inner part of the bar)\footnote{An 
example of the inner part of a bar appearing disc-like can be seen in the simulation shown
  by \citet[][their Fig.~1]{2005ApJ...626..159B}.  If they become
  unstable, they may form the nuclear bar of a double-barred galaxy \citep{1989Natur.338...45S}.} roughly half the length 
of the (full bar)/ansae, NGC~2787 is not the easiest galaxy to model.  
Previously 
modelled by \citet{2003ApJ...597..929E}, the barlens+bar+(large scale disc) was
lumped together there, modelled with an exponential function, and referred to as a pseudobulge.  Within this
`pseudobulge' is an additional bulge component plus a nuclear disc.  
Although an isolated galaxy, it displays a prominent series of nested inner
dust rings, somewhat suggestive of a past accretion event.
\citet{2016ApJS..222...10S}
had some success decomposing the galaxy when they approximated the bar and
ansae \citep[e,g.,][]{2007AJ....134.1863M, 2018ApJ...852..133S} as a
single component. 
\citet{2019ApJ...876..155S}  advanced the decomposition using a
Gaussian ring model for the ansae and a slightly truncated, i.e., double,
exponential disc model.  Such behaviour, where the disc bends down beyond the
bar/ansae, is also observed in simulations \citep{2018ApJ...852..133S}.  The
galaxy components in NGC~2787 are also reminiscent of that seen in NGC~4762
\citep[][their Fig.~3]{2019ApJ...876..155S}.  In Fig.~\ref{Fig-N2787}, we
present what we consider to be a slightly improved fit over that shown in
\citet{2019ApJ...876..155S} for NGC~2787.  Here, the prominence of the
barlens is reduced, no longer forming a dominant bridge between the bar and
ansae.\footnote{We did find a solution with a bulge having a S\'ersic index
  around 2.5; however it also resulted in what was considered an
  unphysical `bridge' pushing the bar and ansae apart.}  From the
equivalent-axis, we find the following spheroid parameters: 
$R_{\rm e,sph,eq}=3\arcsec.9\pm0.4$, 
$\mu_{\rm e,sph}=17.28\pm0.10$ mag arcsec$^{-2}$ (AB) and 
$n_{\rm sph}=0.90\pm0.05$, giving $m_{\rm sph}=11.66\pm0.80$ mag (AB).  
This is $\sim$0.3 mag brighter than reported in \citet{2019ApJ...876..155S}.

\subsubsection{NGC~3377}

\begin{figure}
\begin{center}
\includegraphics[trim=0.0cm 0cm 0.0cm 0cm, width=\columnwidth,
  angle=0]{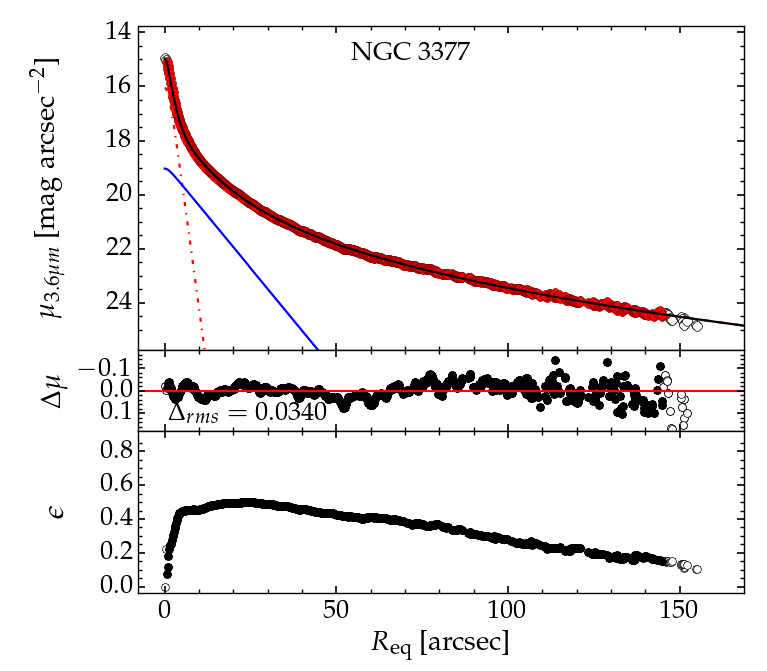}
\caption{Structural decomposition of NGC~3377 using two exponential models for
  the (nuclear and intermediate-scale) discs plus a S\'ersic model for the
  large, dominant spheroid.}
\label{Fig-N3377}
\end{center}
\end{figure}

The ETG NGC~3377 has been remodelled (see Fig.~\ref{Fig-N3377}).  Although
the use of an anti-truncated (up-bending) disc was found to iron the residual
profile flat, we considered this too speculative and opted for the more
traditional components used by \citet{2016ApJ...817...21S}.  Our 3.6~$\mu$m
spheroid parameters from the equivalent-axis are: $R_{\rm
  e,sph,eq}=44\arcsec.2\pm0.6$, $\mu_{\rm e,sph}=21.58\pm0.03$ mag
arcsec$^{-2}$ (AB), 
$n_{\rm sph}=4.48\pm0.08$, and $m_{\rm sph}=9.91\pm0.20$ mag (AB).  In addition to the nuclear disc,
an intermediate-scale disc is required but appears short (given the
ellipticity profile) with a scalelength of just 7$\arcsec$ --- perhaps a clue
that an anti-truncated disc does exist.\footnote{Using an anti-truncated disc
  reduces the spheroid magnitude by just 0.1 mag.}  The galaxy magnitude is
9.76~mag (AB).


%
%



\subsubsection{NGC~4699}

\begin{figure*}
\begin{center}
\includegraphics[trim=0.0cm 0cm 0.0cm 0cm, width=0.45\textwidth, angle=0]{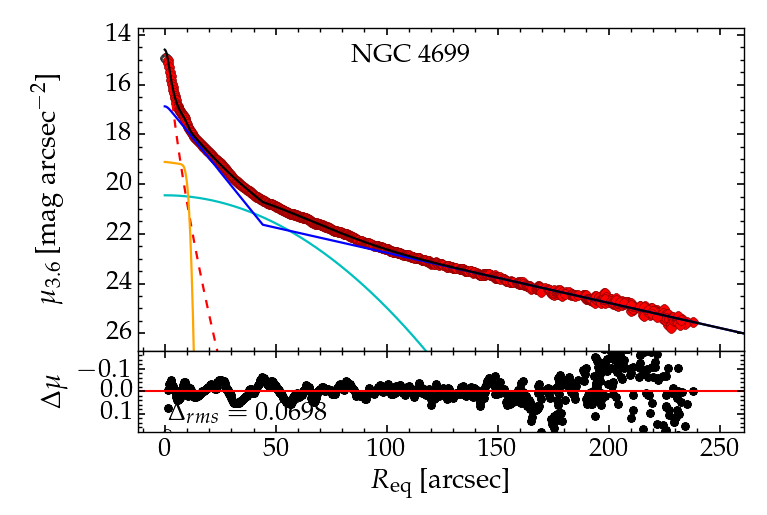}
\includegraphics[trim=0.0cm 0cm 0.0cm 0cm, width=0.45\textwidth, angle=0]{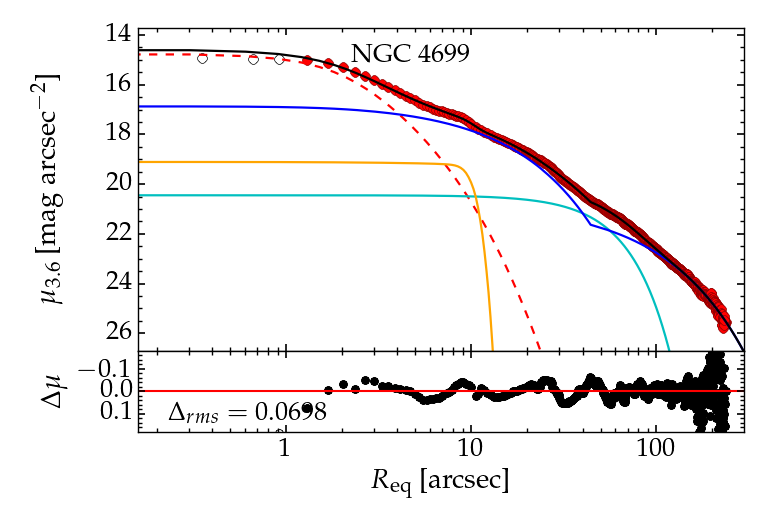}
\caption{Four-component fit: S\'ersic-bulge (red); Ferrers-bar (orange); 
  (anti-truncated exponential)-disc (dark blue); and a broad-Gaussian (cyan) to
  accommodate the spiral arms at $\sim$15--40 and $\sim$60--90$\arcsec$.}
\label{Fig-N4699_240}
\end{center}
\end{figure*}

\begin{figure*}
\begin{center}
\includegraphics[trim=0.0cm 0cm 0.0cm 0cm, width=0.45\textwidth, angle=0]{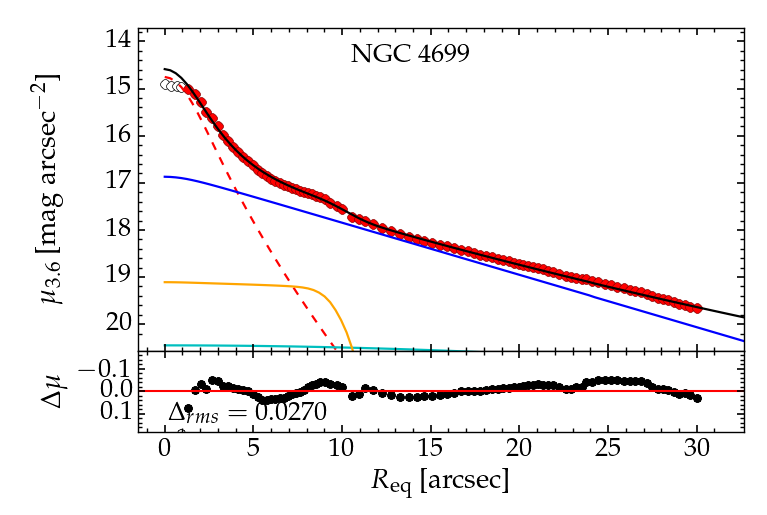}
\includegraphics[trim=0.0cm 0cm 0.0cm 0cm, width=0.45\textwidth, angle=0]{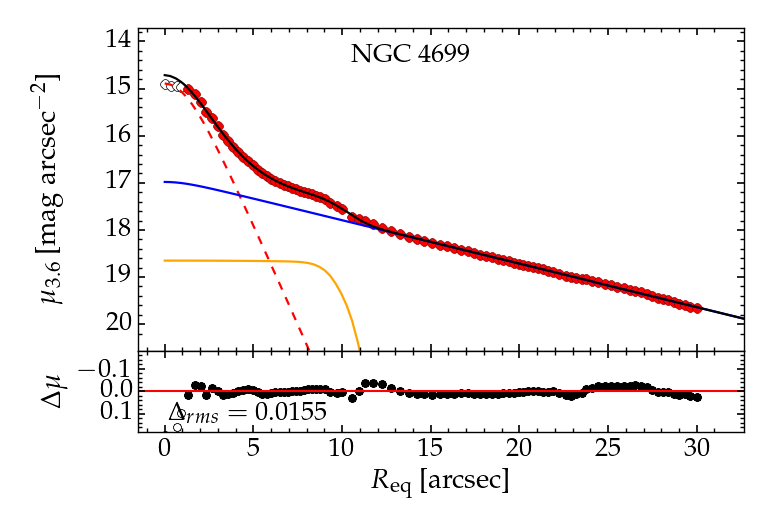}
\caption{NGC~4699. 
Left: Four-component fit from Fig.~\ref{Fig-N4699_240}, zoomed in on the
inner 30$\arcsec$. 
Right: New three-component fit (bulge, bar and disc), without the broad-Gaussian
component, to just the inner 30$\arcsec$.  While this may look more
appropriate, it is misleading due the unmodelled contribution from the spiral
arms at $\sim$15--30 arcseconds.  The stellar mass of the bulge is 0.04 dex
smaller in the 3-component decomposition. 
}
\label{Fig-N4699_30}
\end{center}
\end{figure*}

NGC~4699 is the spiral galaxy with the highest spheroid mass in
\citet{Graham:Sahu:22a}.  However, there are two clues in
\citet{2019ApJ...873...85D} that something may be amiss with their fit, which
treats NGC~4699 like a bulge-dominated system.  The first is that the fitted
ansae --- which captured an inner set of ring-like spirals --- are not at the
end of the bar.  The second clue is that a core-S\'ersic model was used for
the spheroidal component, yet depleted cores are not known to exist in spiral
galaxies, and \citet{2019ApJ...873...85D} remarked that Hubble Space Telescope
imaging showed no core.  We have remodelled the galaxy in
Fig.~\ref{Fig-N4699_240}.\footnote{\citet{2019ApJ...873...85D} also required
  a core-S\'ersic model for the spiral galaxy NGC~3031 (M81), but this was due
  to the saturation of the inner 3$\arcsec$. A core-S\'ersic model was also
  required for NGC~4594 (M104, Sombrero galaxy) due in part to saturation
  within the inner $\sim$2$\arcsec$.}




\citet{2015ApJS..219....4S} and \citet{2015MNRAS.446.4039E} fit the inner data
for NGC~4699 with a S\'ersic function for the bulge plus two exponential 
functions for everything else.  In Fig.~\ref{Fig-N4699_240}, we fit a 
S\'ersic-bulge, a Ferrers-bar, an 
(anti-truncated exponential)-disc plus a broad-Gaussian to capture the bands of
spiral arms.  
The galaxy has two broad sets of ring-like spiral arms at $\sim$15--40 and $\sim$60--90
arcseconds (equivalent-axis).  We found that the simple Gaussian function could
account for both of these, albeit at the expense of introducing an unlikely 
transition region (which we are not interested in) for the anti-truncated disc model. 
The 3.6~$\mu$m bulge magnitude is 11.37 (AB mag).  
%
Using 
(i) a luminosity distance of 23.7~Mpc, 
(ii) $\mathfrak{M}_{3.6} = 6.02$ (AB mag) and 
(iii) $M_*/L_{3.6}=0.66$, based on the (Galactic Extinction)-corrected
colour $B-V = 0.86$ and Equation~4 from \citet{Graham:Sahu:22a}, 
we obtain a stellar mass for the bulge of $\log(M_*/M_\odot) = 10.30\pm0.25$ dex. 
This general procedure was used to obtain the stellar masses used here and
reported in \citet{Graham:Sahu:22a}.  

NGC~4699 is found to have a surprisingly low bulge-to-total stellar mass
ratio of 0.10 given its big stellar mass of $2\times10^{11}$ M$_\odot$. 
Arguably, the inner 
portion of our disc model is too bright with this decomposition, but reducing
this (Fig.~\ref{Fig-N4699_30}, right-hand panel) leads to a brightening
of the bar component, and thus there is little effect on the bulge
magnitude. 
Removing the broad Gaussian, and using three components to fit just the inner
30$\arcsec$, yields a spheroid 
magnitude of 11.46 (AB mag),  
amounting to a small 0.04 dex reduction in the stellar mass that we do not apply.


\subsubsection{NGC~5419}
  
\begin{figure}
\begin{center}
\includegraphics[trim=0.0cm 0cm 0.0cm 0cm, width=\columnwidth, angle=0]{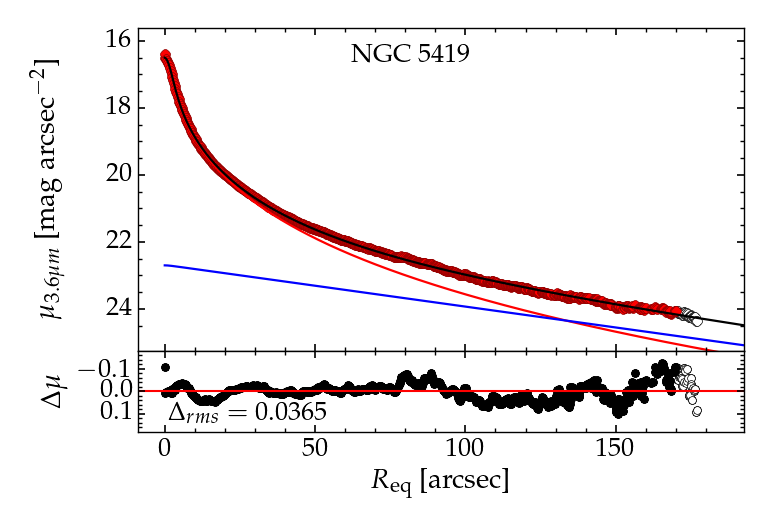}
\caption{Structural decomposition of NGC~5419 using a core-S\'ersic plus 
  exponential function for the spheroid and ICL, respectively.
}
\label{Fig-N5419}
\end{center}
\end{figure}

Following \citet{2019ApJ...876..155S}, we remodel NGC~5419 (Abell S0753) in
Fig.~\ref{Fig-N5419} with a core-S\'ersic function for the spheroid plus an
exponential function for the ICL.  However, we obtain a notably fainter
contribution from the ICL and reduce the `snake-like' pattern in the residual
profile over the (smaller) radial range where the spheroid previously
dominated ($R_{\rm e,sph,eq} \lesssim 40\arcsec$).  The central surface brightness
($\mu_{0,3.6\,\mu m}=22.68$ mag arcsec$^{-2}$) of the exponential model (with
equivalent-axis scalelength $h_{\rm ICL}=87.2\pm1.4$) describing the ICL also
now matches well with that seen around other BCG \citep{2007MNRAS.378.1575S}.
For the spheroid, we find: $R_{\rm e,sph,eq}=38\arcsec.4\pm0.9$, $n_{\rm
  sph}=4.20\pm0.12$ and $m_{\rm sph}=10.00\pm0.55$ mag (AB).

\subsection{Notes on individual galaxies}\label{App_notes}

A table with references to track down all the  measurements of the spheroids' stellar masses and sizes
is given in \citet{Graham:Sahu:22a}.  It also includes the galaxies' stellar
masses and the black hole masses.  Here, we provide additional notes on some
of these individual
galaxies to help fellow researchers follow and appreciate specific changes since KH13.
This builds on the end of Section~\ref{Sec_BCGs} and Section~\ref{Sec_exc}, which
mentioned why we include systems excluded by KH13 

{\bf NGC~221} (M32).  KH13 consider this an elliptical galaxy.  However, it displays a 
bulge/disc light profile \citet{2002ApJ...568L..13G} and has been interpreted
as a heavily stripped S0 galaxy \citep{2001ApJ...557L..39B}.  Furthermore,
its low S\'ersic index bulge has structural parameters consistent with the
location of bulges in the size-mass diagram 
\citet{2013pss6.book...91G}. 
We treat it as a stripped disc galaxy and use the spheroid mass described in
\citet{Graham:Sahu:22a}. 

{\bf NGC~1316, NGC~5128, NGC~2960}.  As noted in Section~\ref{Sec_wet_merger}, 
KH13 consider these elliptical galaxies with overly
bright spheroid magnitudes for their BH mass.  However, they are S0 galaxies,
with NGC~2960 an Sa? galaxy modelled in \citet{2019ApJ...873...85D}, and 
have bulge masses much less than the galaxy masses that KH13 plotted in their
$M_{\rm bh}$-$M_{\rm *,sph}$ diagram (their Fig.~14). 

{\bf NGC 2778}.  KH13 exclude NGC~2778 (which they mislabel in their Fig.~14, assigning the
NGC~2778 label to NGC~4486A) 
because \citet{2011ApJ...729...21S} 
report an upper limit of $M_{\rm bh} = 3\times10^7$ M$_\odot$. 
We have used a distance of 22.1~Mpc and 
$M_{\rm bh} \approx 1.5^{+1.8}_{-0.8}\times10^7$ M$_\odot$ 
\citep{2003ApJ...583...92G}. 
KH13 also considered the galaxy to be an E rather
than an S0 galaxy and consequently over-estimated the spheroid brightness 
by a factor of 4, making it appear somewhat offset in their sample-restricted 
scaling diagram.

{\bf  NGC~3607}.  
KH13 consider NGC~3607 to be an E galaxy; however, it is an ES galaxy with an
embedded intermediate-scale disc \citep{2011MNRAS.414.2923K,
  2016ApJS..222...10S}.  KH13 exclude NGC~3607 due to the potential impact on
the measurement of the BH mass due to dark matter in galaxies with depleted
cores.  However, the strong dust lanes at the centre of NGC~3607 and the dust
rings extending to $\sim$15$\arcsec$ make identifying a depleted core
problematic in this galaxy \citep{2013ApJ...768...36D}.  A somewhat similar
situation, with nuclear dust rings creating a false core, occurred in NGC~4552
\citep{2018MNRAS.478.1161B}.  We elect to retain this galaxy.


{\bf NGC~4261}.   
A new BH mass has been reported for NGC~4261 in \citet{2021ApJ...908...19B},
with $\log(M_{\rm bh}/M_{\odot}) = 9.22\pm0.08$, based on a luminosity
distance of 31.6 Mpc \citep{2001ApJ...546..681T}.  We slightly reduced this
mass by 0.02~dex using an updated distance of 30.4$\pm$2.7 Mpc.  This arises
from a reduction in the distance modulus for NGC~4261 by 0.06~mag after a
recalibration by \citep[][their Section 4.6]{2002MNRAS.330..443B}\footnote{We
  do not use the line from Fig.~7 in \citet{2010ApJ...724..657B}.} and a
further reduction of 0.023~mag arising from a reduced distance modulus for the
Large Magellanic Cloud \citep{2019Natur.567..200P} used by
\citet{2001ApJ...546..681T}.

{\bf NGC~4459}.  
KH13 speculate that the black hole mass in NGC~4459 may be low, having been
measured using ionised gas rotation curves which did not take broad emission
line widths into account.  It appears on the right-hand edge, i.e., possibly
low BH mass, of the distribution of points in the absolute magnitude versus
black hole mass diagram of KH13 (their Fig.~15, see also their Fig.~12)
because they considered it to be an elliptical galaxy rather than a rotating
S0 galaxy \citep{2011MNRAS.414.2923K, 2016ApJS..222...10S}.  Our bulge mass is
a factor of 2 times less than their galaxy mass and is located 
squarely in the midst of things in the $M_{\rm bh}$-$M_{\rm *,sph}$
diagram.  The galaxy also displays no sign of an offset in the $M_{\rm
  bh}$-$\sigma$ diagram, suggesting the BH mass may be fine. 


{\bf NGC~4596}. KH13 advocates excluding this S0 galaxy because it is also
among the set with ionised gas kinematics that did not take broad emission
line widths into account, but it also resides right in the midst of things in
both the $M_{\rm bh}$-$M_{\rm *,sph}$ and $M_{\rm bh}$-$\sigma$ diagrams.
Therefore, as with NGC~4459, its exclusion or inclusion makes no difference.

{\bf NGC~5576}.  
KH13 adjusted the BH mass for NGC~5576, measured by
\citet{2009ApJ...695.1577G}, in an attempt to counter the influence of dark
matter in galaxies with central ``loss cones'' (in phase space) that yield
cores depleted of stars.  However, this is not a core-S\'ersic galaxy; there
is no central deficit of starlight relative to the spheroid's outer light
profile \citep{2012ApJ...755..163D}.  The increased BH mass in KH13 is
$\sim$50 per cent larger but also has an increased uncertainty range keeping
it consistent with the original mass we use.  All elliptical galaxies like
NGC~5576 are, however, built by mergers, as Fig.~3 in
\citet{2009AJ....138.1417T} shows for NGC~5576.

{\bf NGC 7052}.  This galaxy has a new BH mass in \citet{2021MNRAS.503.5984S}, based on
molecular gas, which is an order of
magnitude larger, at $\log(M_{\rm bh}/M_{\odot}) = 9.38\pm0.05$ (D=66.4~Mpc), 
than the previous estimate based on the ionised gas
\citep{1998AJ....116.2220V}.  It highlights how the small errors reported on
black hole masses only capture the statistical error rather than the likely
more dominant systematic error due to, for example, the non-gravitational motion
of the gas.  Based on our preferred distance of 61.9 Mpc
\citep{2021ApJS..255...21J}, this yields $\log(M_{\rm bh}/M_{\odot}) = 9.35\pm0.05$.

\bsp    
\label{lastpage}
\end{document}